\documentclass[]{spie}  
\usepackage[]{graphicx}

\newcommand{\msun}{M_\odot}

\title{AXIS: A Probe Class Next Generation High Angular Resolution X-ray Imaging Satellite} 

\author{R. Mushotzky for the AXIS Team
\skiplinehalf
\supit{a}University of Maryland, Department of Astronomy, College Park, Maryland; \\
}

\authorinfo{Further author information: (Send correspondence to R.M.)\\R.M.: E-mail: richard@astro.umd.edu, Telephone: 1-301-405-6853\\}

\begin{document} 
  \maketitle 

\begin{abstract}
AXIS is a probe-class concept under study for submission to the 2020 Decadal survey. AXIS will extend and enhance the science of high angular resolution X-ray imaging and spectroscopy in the next decade with $\sim 0.4''$ angular resolution over a 24$' \times 24'$ field of view, with 0.3$"$ in the central 14$' \times 14'$, and an order of magnitude more collecting area than Chandra in the 0.3$-$12 keV band with a cost consistent with a probe.

These capabilities are made possible by precision-polished lightweight single-crystal silicon optics achieving both high angular resolution and large collecting area, and next generation small-pixel silicon detectors adequately sampling the point spread function and allowing timing science and preventing pile up with high read-out rate.  We have selected a low earth orbit to enable rapid target of opportunity response, similar to Swift, with a high observing efficiency, low detector background and long detector life. 

The combination opens a wide variety of new and exciting science such as: (1) measuring the event horizon scale structure in AGN accretion disks and the spins of supermassive black holes through observations of gravitationally-microlensed quasars; (ii) determining AGN and starburst feedback in galaxies and galaxy clusters through direct imaging of winds and interaction of jets and via spatially resolved imaging of galaxies at high-z; (iii) fueling of AGN by probing the Bondi radius of over 20 nearby galaxies; (iv) hierarchical structure formation and the SMBH merger rate through measurement of the occurrence rate of dual AGN and occupation fraction of SMBHs; (v) advancing SNR physics and galaxy ecology through large detailed samples of SNR in nearby galaxies; (vi) measuring the Cosmic Web through its connection to cluster outskirts; (vii) a wide variety of time domain science including rapid response to targets of opportunity. 
  
With a nominal 2028 launch, AXIS benefits from natural synergies with the ELTs, LSST, ALMA, WFIRST and ATHENA. The AXIS team welcomes input and feedback from the community in preparation for the 2020 Decadal review.
  
\end{abstract}



\section{INTRODUCTION}
\label{sec:intro}  

The Advanced X-ray Imaging Satellite (AXIS) will be the premier high angular resolution X-ray mission for the late-2020s. With sub-arcsecond (0.4$''$ or better) angular resolution across a large field of view (24$' \times 24'$) and a collecting area an order of magnitude larger than Chandra, this mission represents a huge leap in our ability to study the hot and energetic universe. AXIS will provide a fundamentally new window on the growth modes and energy generation of supermassive black holes over cosmic time through measurements of black hole spin in quasars that are strongly gravitationally lensed.  Building on the seminal Chandra-results on nearby galaxy clusters, AXIS will provide our first view of the physics of active galactic nuclei (AGN) feedback in high-redshift massive galaxy clusters as well as low-mass groups and galaxies, allowing a full picture of these feedback processes to be constructed. AXIS will probe the processes that determine AGN fueling via imaging spectroscopy within the black hole sphere of influence of many nearby massive galaxies. These are just three examples of the high-resolution X-ray science that will be underpinned by AXIS a decade from now as it takes its place alongside the next generation of astronomical observatories (ATHENA, JWST, WFIRST, LSST, SKA, ALMA, TMT, ELT, CTA).

   \begin{figure}[htb]
   \begin{center}
   \begin{tabular}{c}
   \includegraphics[height=7cm]{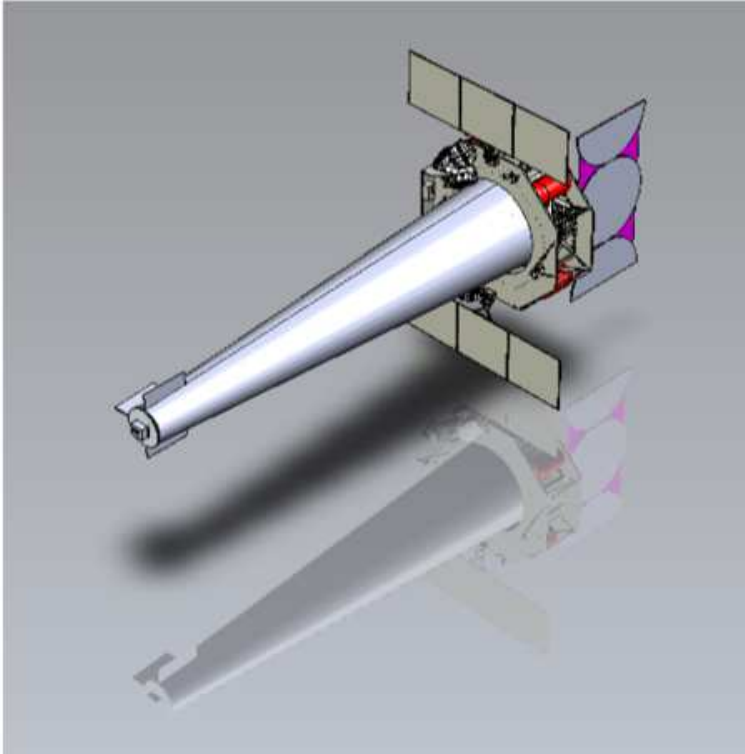}
   \end{tabular}
   \end{center}
   \caption[example] 
   { \label{mockup} 
CAD drawing of the AXIS spacecraft from the GSFC MDL study}
   \end{figure} 

We will build on the mirror technology development initiated by the Constellation-X/IXO program and sustained by NASA's Strategic Astrophysics Technology program, whose goal was to produce high-resolution and lightweight X-ray optics at reasonable cost. The technology is based on  precision polishing and light weighting of single-crystal silicon, leveraging knowledge and equipment that have been developed by the semiconductor industry.  At present this technology has achieved 2.2$''$ angular resolution for a mirror pair \cite{Zhang18}, with a clear path toward much better angular resolution in the early 2020s.  The technology builds on recent developments in the semiconductor industry: (i.) the inexpensive and abundant availability of large blocks of single crystal silicon; and (ii.) revolutionary advances in rapid, deterministic, precision polishing of wafers. The baseline detector is similar to the Chandra CCD but benefits from 25 years of technology development, allowing the sampling of the PSF, in turn producing higher effective angular resolution, faster readout time and broader bandpass. CCD and CMOS detectors with the needed properties are being developed today \cite{Falcone18, Bautz18}.

AXIS has been studied at the NASA Goddard Space Flight Center (GSFC) Instrument and Mission Design Labs (IDL/MDL) and a spacecraft design has been developed using proven spacecraft components and methods for attitude reconstruction  compatible with the angular resolution. The estimated mission cost was well below the \$1B cost cap for a Probe mission in 2018 dollars. The single technology area needing major development is the construction of the X-ray mirror.  Since the mirror needed for this mission is a smaller version of the one envisioned for X-ray Surveyor (XRS), the necessary mirror technology development is already underway and successful development of AXIS will pave the way for the larger XRS mirror.

The AXIS science objectives are directly responsive to the goals set forth by the Astro2010 Decadal Survey and the NASA Astrophysics Roadmap. Some of our proposed science highlights include: probing the regions within 20 Schwarzschild radii of accreting black holes via gravitational lensing, determining the rate of black hole mergers by measuring the space density and nature of the hosts of dual AGN, understanding why most black holes in the local universe are not active by imaging the Bondi accretion radius, measuring feedback and star formation by X-ray imaging of normal galaxies to z$>$2, vastly enlarging the sample of high signal-to-noise high angular resolution images of supernova remnants, probing the structure of clusters of galaxies from the very central regions to beyond the virial radius and opening up a new window of time domain astrophysics. 

\begin{table}[h]
\label{tab:fonts}
\begin{center}       
\begin{tabular}{|l|l|} 
\hline
\rule[-1ex]{0pt}{3.5ex}  Angular resolution & 0.4$"$ over 24$' \times 24'$, 0.3$"$ over central 14$' \times 14'$ \\
\hline
\rule[-1ex]{0pt}{3.5ex}  Bandpass & 0.1$-$10 keV   \\
\hline
\rule[-1ex]{0pt}{3.5ex}  Effective area & 7000 cm @ 1 keV; 1000 cm @ 6 keV  \\
\hline
\rule[-1ex]{0pt}{3.5ex}  Energy resolution & 150 eV @ 6 keV (CCD resolution)  \\
\hline
\rule[-1ex]{0pt}{3.5ex}  Total count rate per source & $\sim 10$x Chandra at time of Chandra launch  \\
\hline
\rule[-1ex]{0pt}{3.5ex}  Internal background rate & 10-20x less than Chandra depending on energy \\
\hline
\rule[-1ex]{0pt}{3.5ex}  Diffuse signal to background ratio & 40$-$60$\times$ lower than Chandra \\
\hline
\end{tabular}
\caption{Basic parameters of AXIS} 
\end{center}
\end{table} 

   \begin{figure}[htb]
   \begin{center}
   \begin{tabular}{c}
   \includegraphics[height=8cm]{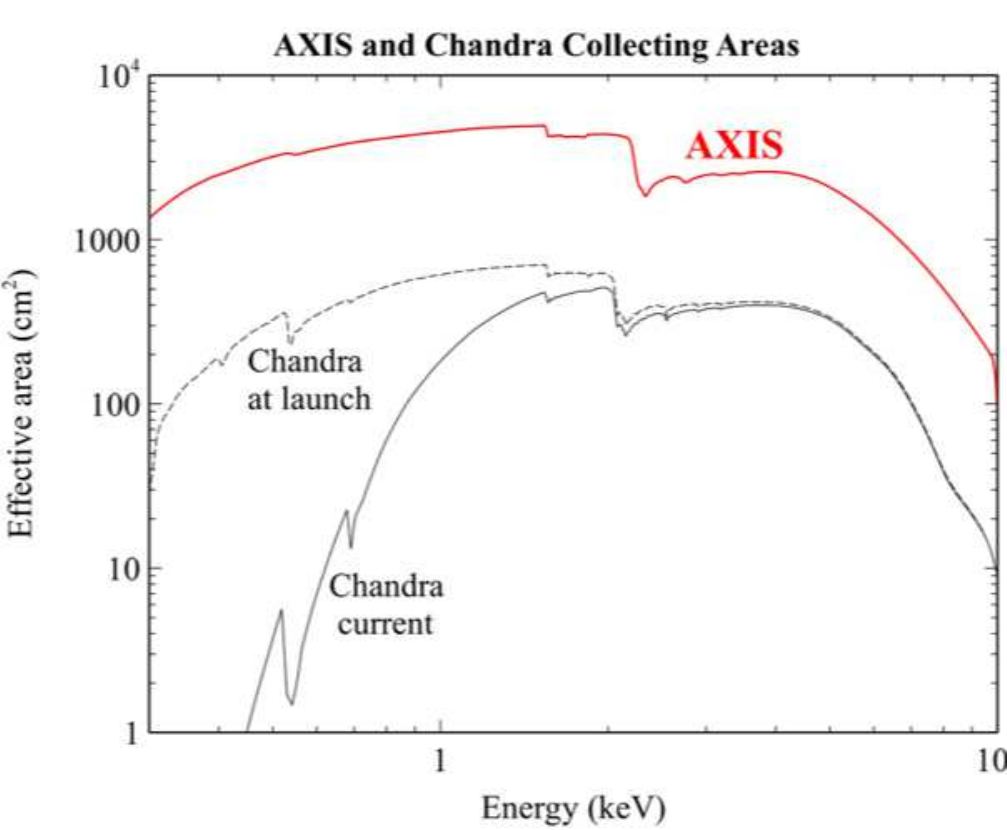}
   \end{tabular}
   \end{center}
   \caption[example] 
   { \label{collecting_area} 
Comparison of the AXIS collecting area including detector efficiency and filters to that of Chandra at launch. The lower line shows the effect of the growth of contamination on the Chanda ACIS detector.}
   \end{figure} 

   \begin{figure}[htb]
   \begin{center}
   \begin{tabular}{c}
   \includegraphics[height=9cm]{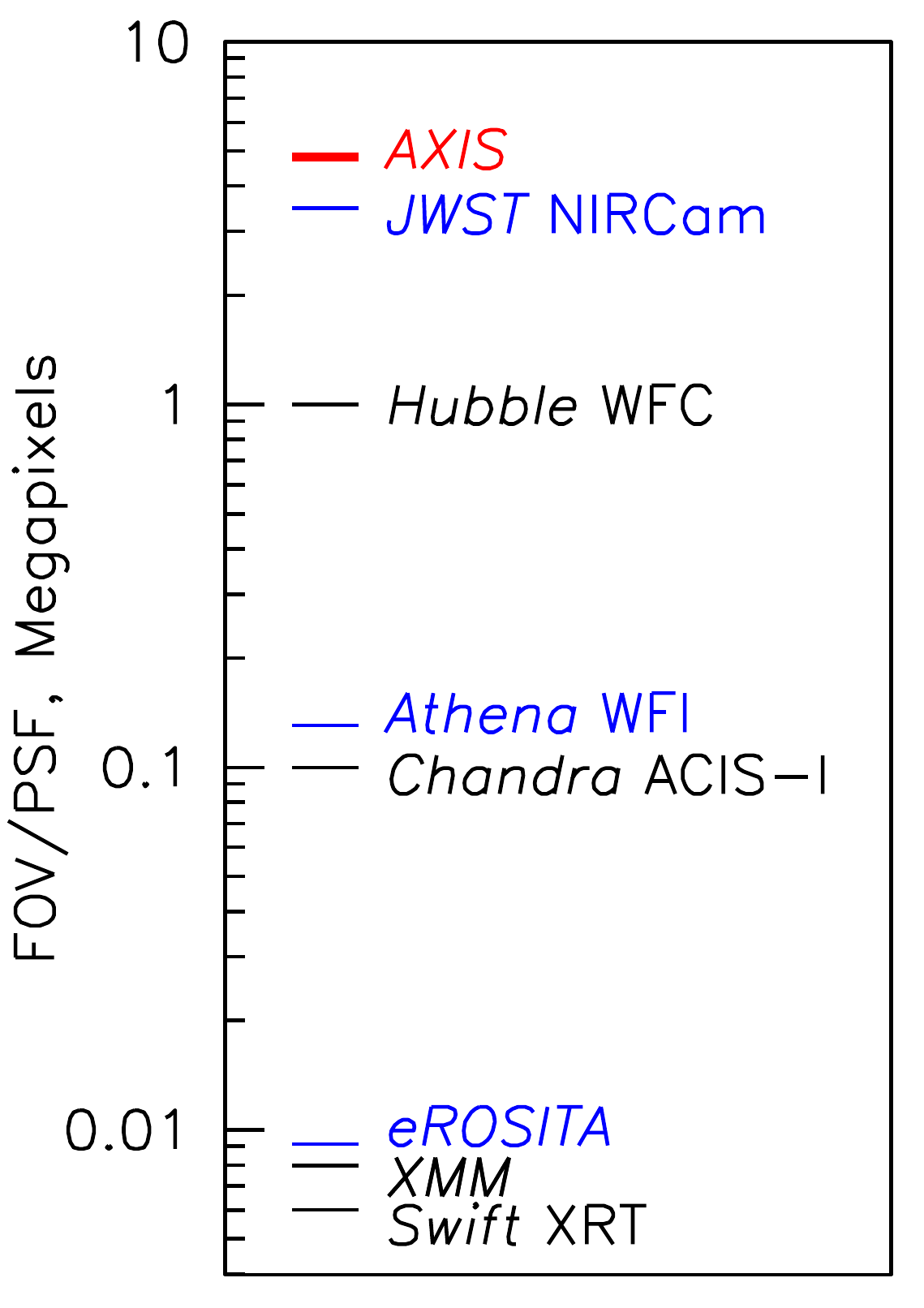}
   \includegraphics[height=9cm]{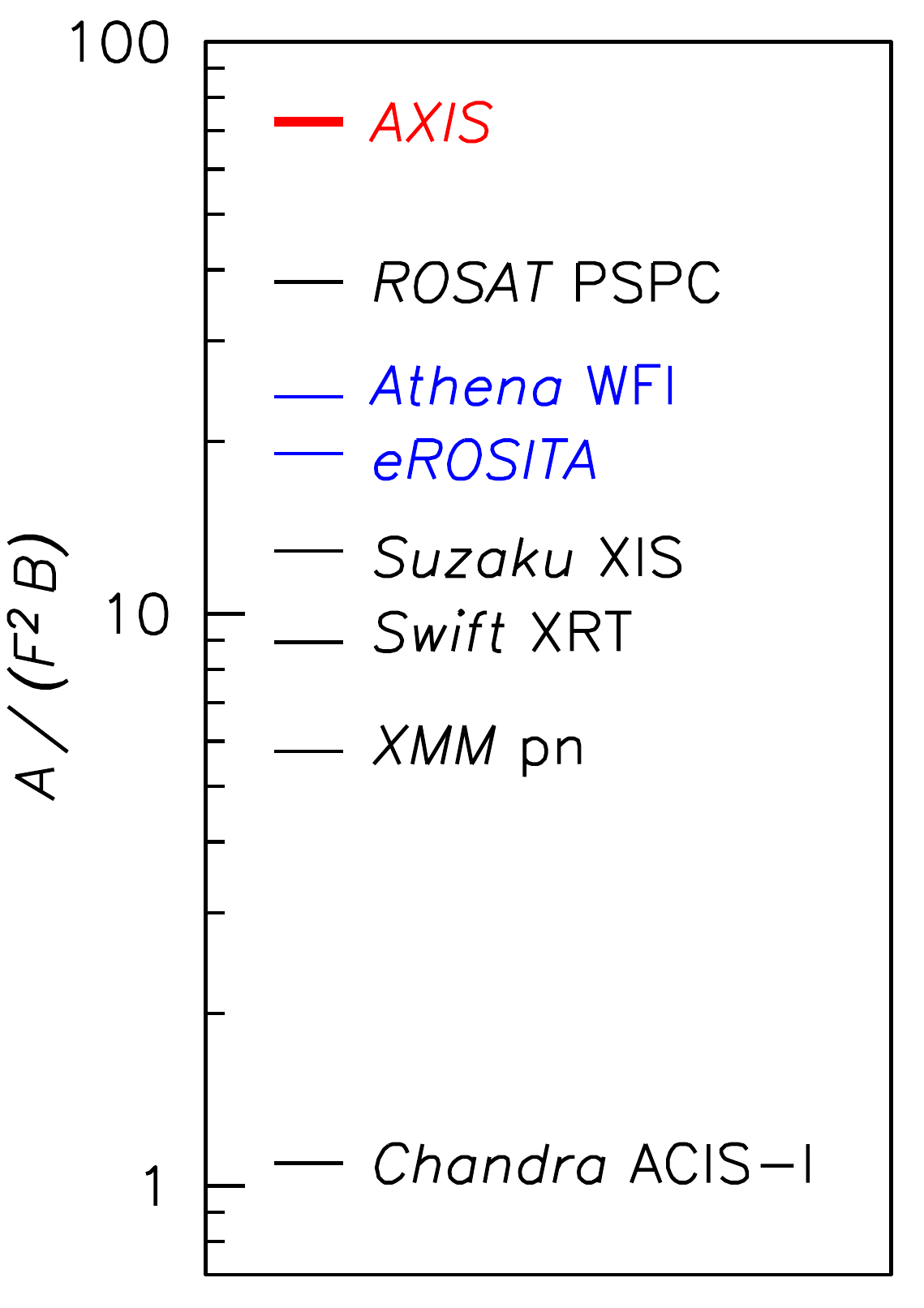}
   \end{tabular}
   \end{center}
   \caption[example] 
   { \label{mockup} 
Two figures of merit for AXIS. Left: The amount of
imaging information in each pointing, given by the number of effective
resolution elements (PSF HPD) per field of view, taking into account
the worsening of the PSF with off-axis angle. Right: Sensitivity to
faint extended emission, given by the ratio of the surface brightness
of a celestial source to that of the detector background. A is
effective area at E~1 keV, F is focal length, B is approximate
detector background per unit detector area (using for simplicity B=1
for missions at low-earth orbits and B=4 for high orbits)}
   \end{figure} 

\section{AXIS SCIENCE}
\label{science}

The need for high angular resolution (sub-arcsec) in astrophysics is evident across the entire electromagnetic spectrum from sub-arcsec imaging and spectroscopy in the radio and millimeter bands (VLBI, ALMA), to the infrared (JWST), optical (HST) and X-ray (Chandra). It is essential for resolving the critical physical scales of virtually all classes of objects and the desire to extend such studies to the highest redshift.  These goals have driven the development of virtually all new telescopes in the radio (EHT, SKA), infrared and optical (interferometers, coronagraphs and adaptive optics on large ground based telescopes and WFIRST), but so far there has been no development of new high-resolution telescopes in the X-ray band.  Astronomy is a multi-wavelength science with almost every class of astrophysical object having critical diagnostic information available only with appropriate multi-wavelength data.  

We have baselined the lowest cost mission that meets the science case and does not exceed the Probe cost cap. As shown by the 2012 NASA-sponsored X-ray Mission Concepts Study it is not possible to develop a high resolution X-ray imaging mission in an Explorer budget. We have designed AXIS to fit within the probe cost range, trying to obtain the best possible high angular resolution science in the X-ray band while foregoing additional capabilities. AXIS will also be a crucial step towards a strategic class mission such as Lynx that would achieve the trifecta of high-throughput, high-resolution imaging and high-energy-resolution spectroscopy. 

AXIS will be a giant step in angular resolution, sensitivity, lower background, quick response time and much better soft response than anything ever built before, allowing one to push the limits of astrophysical research into new and unexplored areas.

\subsection{Galaxy Formation and Evolution}

The formation and evolution of galaxies is a keystone topic of modern astrophysics research, which involves characterizing stellar populations and understanding the processes that determine them and changes over cosmic time. AXIS will make transformative contributions in both areas. 

\subsubsection{Tracking Galaxy Growth Across Cosmic Time}

Hard X-rays are a proxy for the star-formation rate (SFR) through the total emission of high-mass X-ray binaries (e.g., SFR $\approx$ L$_{X, 2-10}$ keV / 5$\times$10$^{39}$ erg s$^{-1}$, \cite{Mineo14}). X-rays above 2 keV (rest-frame) are insensitive to reddening and have completely different systematics than other star formation rate indicators, so they are an excellent probe of SFR up to z $<3-4$, which is the period in which almost all of the stars in the Universe were formed. 

Spatially-resolved observations are crucial to determining how galaxies build up stellar populations (e.g., inside-out star-formation or star formation within outflows), as well as to isolate X-rays from star formation from AGN. Recent HST observations show that, over a wide range in redshift, "normal" galaxies have characteristic sizes between 2.5$-$4.5 kpc \cite{Ferguson04}, which, for z$>$1, is smaller than 1 arcsec. This is resolvable with the AXIS resolution of $<0.5''$. At a sensitivity of 10$^{-17}$ erg s$^{-1}$ cm$^{-2}$ in 1 Ms at E$<$1 keV, AXIS will detect normal star-forming galaxies at z $\sim 2$ and lensed rapidly star forming galaxies to z $\sim 4$, which allows measurement of the temperature and abundance of the ISM.  

The hot gas associated with star formation has a temperature between 3$-$10 $\times$10$^{6}$ K, and the primary emission lines (OVII 0.57 keV, OVIII 0.65 keV, and various Fe L-shell lines below E=1.5 keV) are measurable with AXIS out to z $\sim$ 4.  

\begin{figure}[htb]
   \begin{center}
   \begin{tabular}{c}
   \includegraphics[height=6cm]{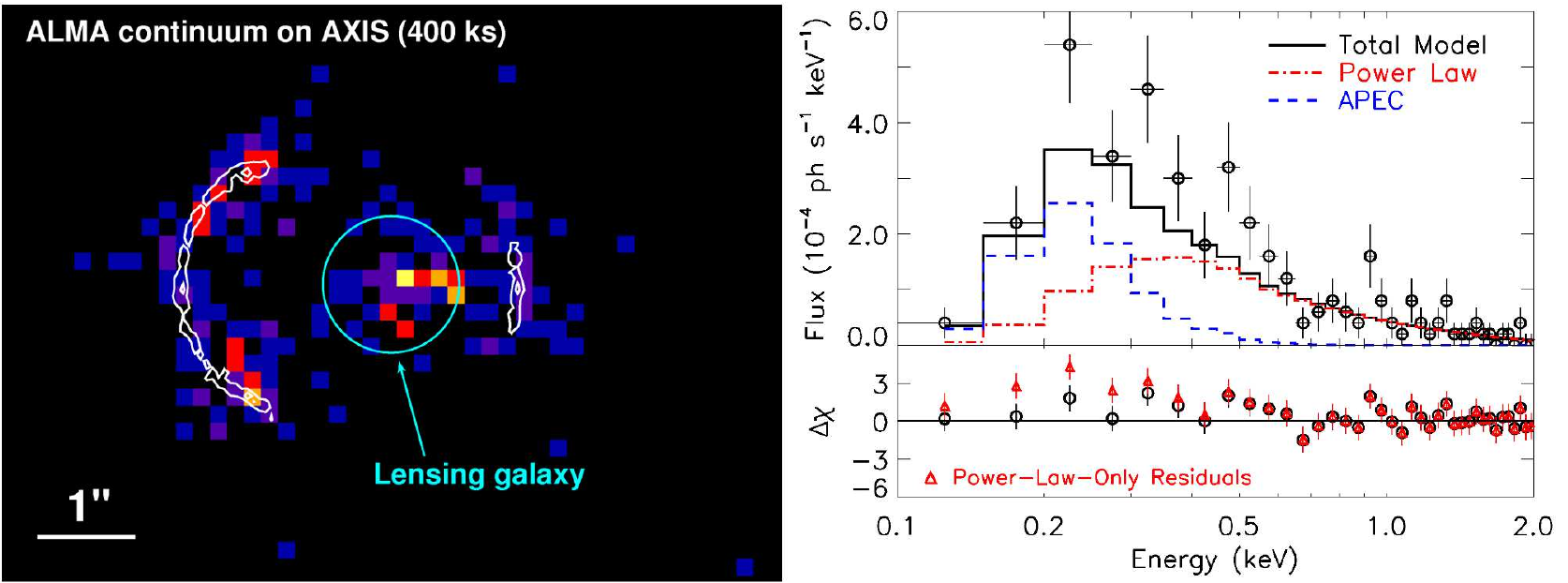}
   \end{tabular}
   \end{center}
   \caption[example] 
   { \label{alma_contours_sdp81} 
ALMA continuum contours overlaid on a 400 ks AXIS simulation of the strongly lensed galaxy SDP-81 at z=3.042 \cite{Dye15}. The lensing galaxy (at z=0.3) is also visible. The right panel shows the spectrum from the lensed galaxy. The X-ray emission comes from the sum of high-mass X-ray binaries and  hot gas.}
   \end{figure}

Observations of strongly lensed, very high redshift galaxies, discovered by Herschel and observed with ALMA \cite{Negrello17}, allows AXIS to obtain spatially resolved measurements in 500ks out to redshifts of $\sim 4$. Figure~\ref{alma_contours_sdp81} shows that AXIS can accumulate enough photons to measure the X-ray luminosity in several regions for a lensed galaxy at z=3.042 with SFR $\approx$ 500 $\msun$ yr$^{-1}$ \cite{Dye15}, and can detect the hot phase of the ISM.

At lower redshift, the combination of high sensitivity and spatial resolution enables the detection of X-ray binaries down to L$_{X, 2-10}$ keV $< 10^{38}$ erg s$^{-1}$ out to about 200 Mpc in moderately deep images, and down to L$_{X} <10^{35}$ erg s$^{-1}$ for a large sample of local galaxies. The brightest sources (ULXs) will be easily detected out to z $<$ 0.5. This will enable a calibration of the binary luminosity-stellar mass relation, provide independent measures of the initial-mass function, and constrain binary formation channels. 

\subsubsection{Galactic Feedback in Action}

A galaxy's stellar population is determined by the competition between gravity and non-gravitational "feedback" processes. These processes, including supernovae and AGN winds, are energetically dominated by hot gas (T $> 10^{6}$ K), which is only visible in X-rays, occur in small regions, requiring high-resolution X-ray observations. AXIS will show how galaxies expel gas and also prevent gas from cooling. 

\begin{figure}[htb]
   \begin{center}
   \begin{tabular}{c}
   \includegraphics[height=6cm]{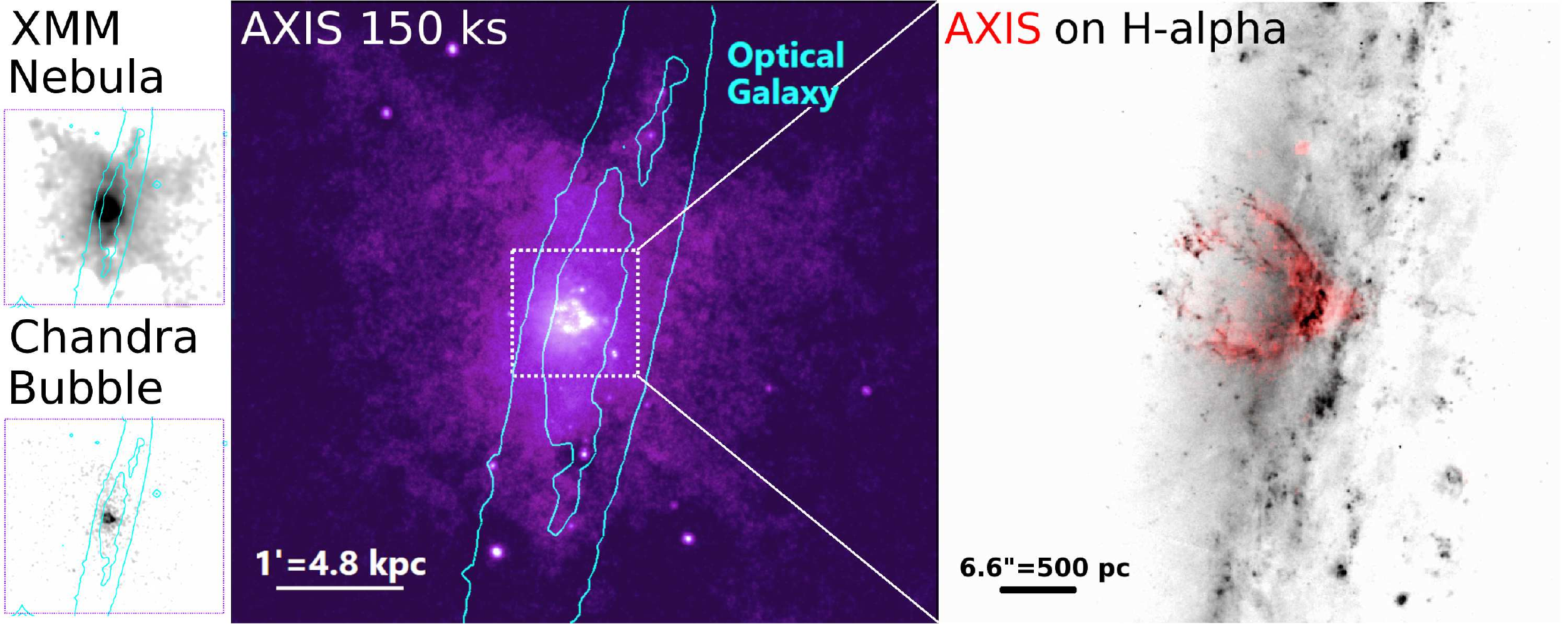}
   \end{tabular}
   \end{center}
   \caption[example] 
   { \label{ngc3079} 
A 150 ks AXIS simulation of the galactic wind in NGC 3079, a nearby, edge-on spiral galaxy, based on existing XMM-Newton and Chandra data (left). This exposure resolves the nebula into a network of thin filaments more than 20 kpc off-plane, enabling pressure measurements to determine the origin of the soft X-rays and obtain an accurate energy budget for the wind. At smaller scales, the knots in the bright super-bubble region of the galaxy (right, overlaid on an HST image) are resolved on $<$100 pc scales. AXIS will determine accurate temperatures, metallicities, and densities.}
   \end{figure}

During periods of intense star formation, superheated gas is formed, whose pressure drives a wind out of the galaxy at (up to) thousands of km s$^{-1}$ \cite{Veilleux05}. These winds may be the major agents of galactic feedback, especially at z$>$1. AXIS will transform our understanding of stellar feedback through mapping the temperature, density and chemical composition of nearby winds at high spatial resolution (Figure~\ref{ngc3079}), complementary to high resolution spectroscopy with Athena.

AXIS will detect galactic winds at z$>$1, via measurement of Fe-L lines redshifted to 0.3$-$0.5 keV, on scales of 5$-$20 kpc. These observations will determine the enrichment of the circumgalactic and intergalactic medium, and will show whether such winds can quench star formation. 

\begin{figure}[htb]
   \begin{center}
   \begin{tabular}{c}
   \includegraphics[height=6cm]{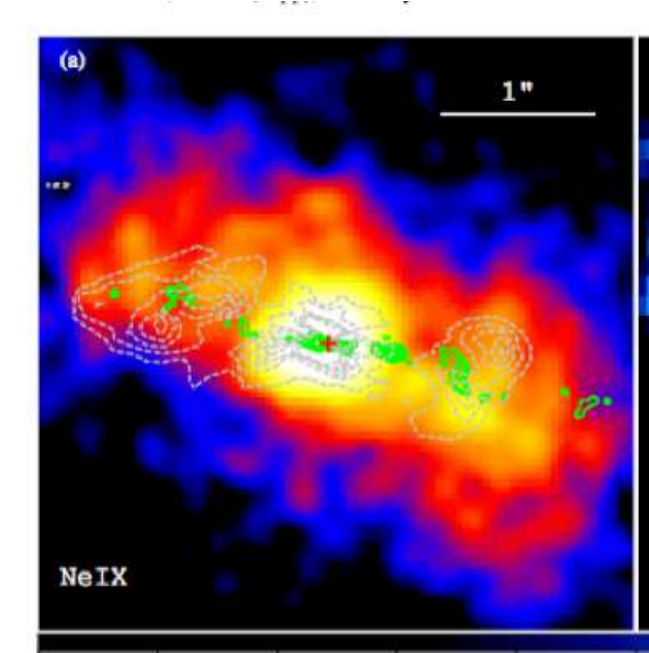}
   \end{tabular}
   \end{center}
   \caption[example] 
   { \label{ngc4151} 
Chandra Ne IX (0.9 keV) image of the central region of the active galaxy NGC4151 with radio contours (green) and [Fe II] contours that show the presence of ionized gas on scales $<2''$ \cite{Wang11}. The hot gas follows the jet, indicating direct heating of the gas by the jet. In other nearby systems, photoionized X-ray emitting gas directly tracks line-driven winds. Both cases show the immediate impact of feedback.}
   \end{figure}

Chandra images of a few AGN (e.g., NGC 4151 \cite{Wang11}; Figure~\ref{ngc4151}) show hot gas in the central $\sim 100$ pc. This gas is either heated by interaction with a jet or is a fast photoionized wind interacting with the ISM \cite{Tombesi15}.  The X-ray images provide the best opportunity to catch feedback in action, as winds and jets are accelerated close to the black hole and determine the physics of the wind, e.g. whether it is energy or momentum driven. Chandra has sufficient resolution, but a poor soft response and under-sampled PSF that makes further progress difficult. 

AXIS will make detailed maps with precise temperature and density measurements, to determine the heating rate on scales smaller than the narrow-line region for a sample covering a range of Eddington ratios and SFR. The high sensitivity and angular resolution of AXIS allow determination of  the spatial connection between the main different phases of AGN winds, i.e., the putative "hot phase" with the colder, neutral and molecular phases.  

High-resolution X-ray images of galaxy clusters revolutionized AGN and cluster science by demonstrating that AGNs prevent most of the surrounding hot gas from cooling \cite{Fabian12}. AGNs play a similar role in galaxies, especially elliptical galaxies. Like clusters, these galaxies have hot atmospheres that capture the AGN energy, but these atmospheres are fainter, cooler, and smaller than in clusters. The AXIS band covers the diagnostic  lines from C, N, O, Ne, Mg and Fe. With its superior low-energy response, AXIS will do for individual galaxies what Chandra did for galaxy clusters: identify many shocks and jet-blown cavities, map the temperature, metallicity, and entropy on sub-kpc scales, and measure the direct impact of the AGN jet and wind feedback on the ISM.

\subsection{Black Holes}

\subsubsection{Gravitational Lensing}

\begin{figure}[htb]
   \begin{center}
   \begin{tabular}{c}
   \includegraphics[height=5cm]{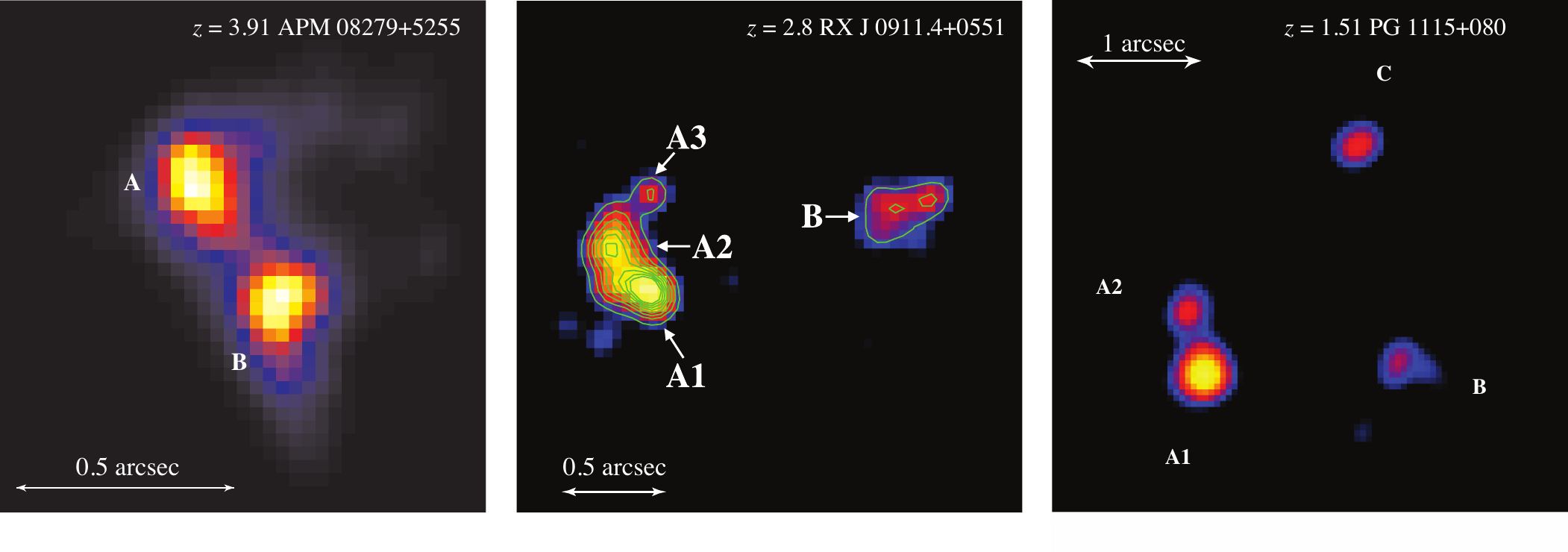}
   \end{tabular}
   \end{center}
   \caption[example] 
   { \label{lensed_quasars} 
Lucy-Richardson deconvolved Chandra images of the gravitationally lensed quasars APM 08279$+$5255 (left) RX J0911.4$+$0551 (middle) and  PG 1115$+$080 (right).}
   \end{figure}

Chandra has enabled breakthroughs in the study of gravitational lensing of supermassive black holes (eg., see Figure~\ref{lensed_quasars}). If the geometry is just right, gravitational lensing by a foreground galaxy can produce several images of a background quasar, separated by a few arcseconds or less.  Each image then flickers due to the action of microlensing by individual stars within the lensing galaxy, and a statistical analysis of this flicker reveals information about the micro-arcsecond scale structures of the inner accretion disk and the hot corona and thus in a sense map the central tens of Schwarzschild radii near the black hole.  Analysis of observations of $\sim 10$ strongly lensed quasars, which are at the limit of Chandra's present day sensitivity \cite{Pooley12, Chartas09, Chartas12, Chartas16}, shows that the X-ray emission originates within 20 Schwarzschild radii and have provided coarse maps of the Fe K$\alpha$ and X-ray continuum emitting regions \cite{Morgan10, Pooley09, Chartas16}.  These spectacular results barely touch the power of x-ray lensing to reveal structures extremely close to the black hole. 

The Chandra sample is severely limited to only a  few quasars that are X-ray bright enough and with image separations large enough to produce good S/N spectra and images.  AXIS will increase the sample size by $\sim 2$ orders of magnitude and the signal-to-noise ratio in each object.  It will directly image the central regions near a black hole, which, when combined with the Athena high signal-to-noise, high-resolution spectra of non-lensed sources will produce a major breakthrough in our understanding of these extreme regions.

\begin{figure}[htb]
   \begin{center}
   \begin{tabular}{c}
   \includegraphics[height=6cm]{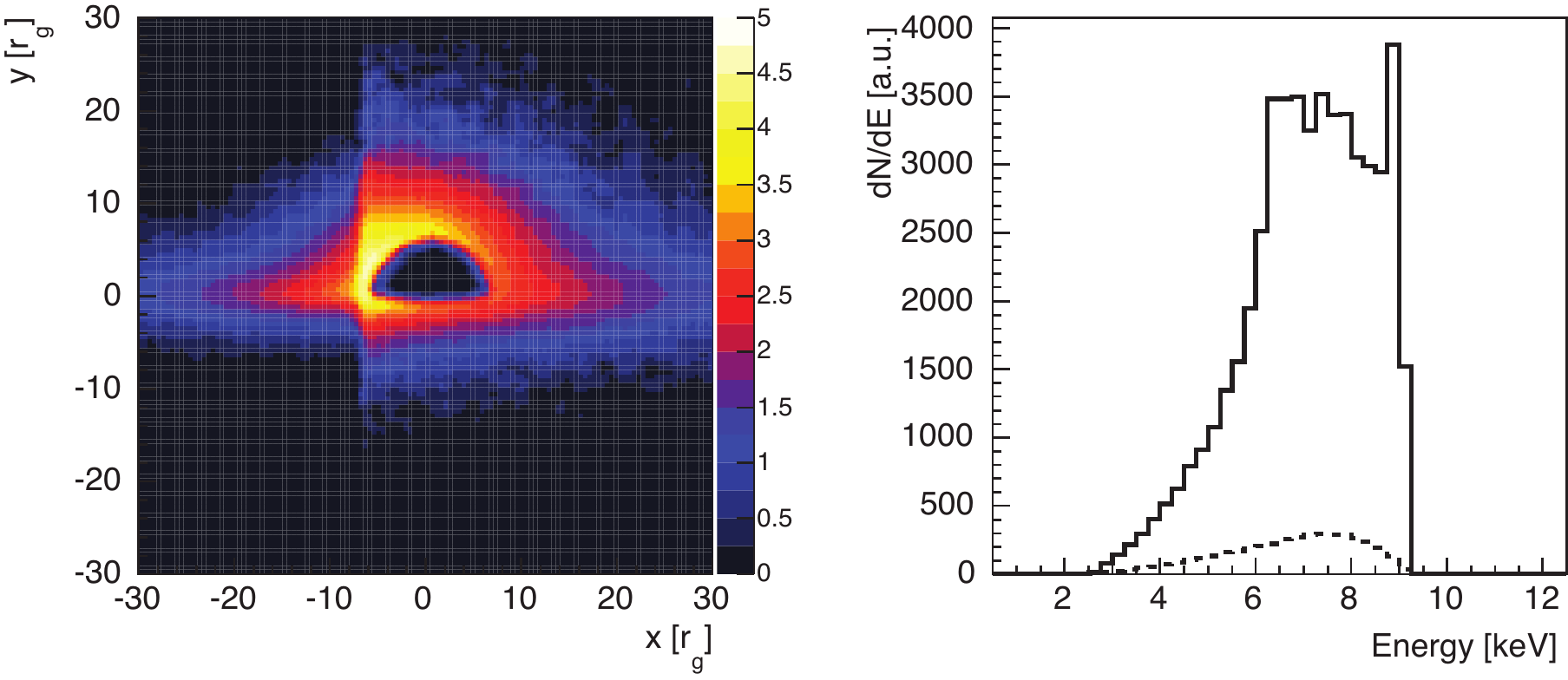}
   \end{tabular}
   \end{center}
   \caption[example] 
   { \label{microlensing_FeKa} 
The left panel shows a simulated image of the surface brightness of the Fe K$\alpha$ line emission during a caustic crossing The surface brightness exhibits a left-right asymmetry owing to the motion of the accretion disk plasma toward (left) or away from (right) the observer. The right panel shows the resulting energy spectrum of the Fe K$\alpha$ emission in the rest frame of the source during this moment.  The spectrum and flux change radically during the crossing revealing the structure of the central regions The dashed line shows the Fe K$\alpha$ line with no microlensing present.}
   \end{figure}

The very small angular size of the X-ray source gives rise to the largest amplitude microlensing signals and thus place the strongest constraints on the mean surface density of the lensing galaxy, the dark matter fraction in the lensing galaxies, the mean surface density in the stars of the lensing galaxy, and the mean mass $<$M$>$ of the stars in the lens.  As shown in \cite{Pooley09}, the small number of Chandra observations constrains the ratio of stellar matter to dark matter.  AXIS's higher signal-to-noise data and a larger sample can drastically improve these constraints. During a caustic crossing the flux and spectrum of the source is radically changed (Figure~\ref{microlensing_FeKa})

Recent Chandra observations of several X-ray bright lensed quasars have constrained the inclination angle, size of the ISCO, and spin of quasars \cite{Chartas17, Krawczynski17, Chartas17} of distant quasars. AXIS will extend such results to a large sample using the  LSST predicted sample of 4000 lensed quasars.  In addition,  the critical AXIS observations of a caustic crossing can be triggered by the LSST monitoring of lensed AGN. This sample will  map the evolution of black hole spin between redshifts of  about $\sim$ 4 and $\sim$ 0.5 \cite{Volonteri13} and  the  accretion disk structure as a function of AGN luminosity, black hole mass, redshift, and host galaxy properties, achieving a fundamental breakthrough in our understanding of accretion disks, the hot X-ray emitting corona, the mean surface density of the lensing galaxy, the dark matter fraction in the lensing galaxies, the mean surface density in the stars of the lensing galaxy, the mean mass of the stars in the lens, and the geometry of the Fe K emission region.  

The high magnification of  the source during the lensing event will produce much higher signal to noise spectra which can \cite{Reis14, Walton16},  combined with the AXIS rapid response to targets of opportunity, detect the presence of extremely fast outflows at high redshift in an unbiased sample.  AXIS will thus drastically increase the number of objects with spectroscopy of the outflows at high redshift where theory indicates that feedback is dominant.

\subsubsection{The Sphere of Influence of Supermassive Black Holes}

A supermassive black hole sitting in the hot X-ray emitting atmosphere of a giant elliptical galaxy will accrete a nearly spherical inflow of hot gas, via "Bondi accretion" \cite{Baganoff03}.  The temperature and density of the gravitationally captured, inflowing material are expected to increase as it approaches the black hole in a simple, predictable fashion, producing a deterministic accretion rate onto the black hole.  For the central density and temperatures seen in many nearby elliptical galaxies and Sgr A*, the observed luminosities are many orders of magnitude smaller than predictions based on the Bondi accretion rate assuming a standard 10\% radiative efficiency.  Either the radiative efficiency is significantly overestimated or considerably less gas is accreted than predicted.  This discrepancy highlights a major gap in our understanding of accretion and the quiescence of the local black hole population, and is important for understanding the origin and evolution of black holes across cosmic time.

   \begin{figure}[htb]
   \begin{center}
   \begin{tabular}{c}
   \includegraphics[height=7cm]{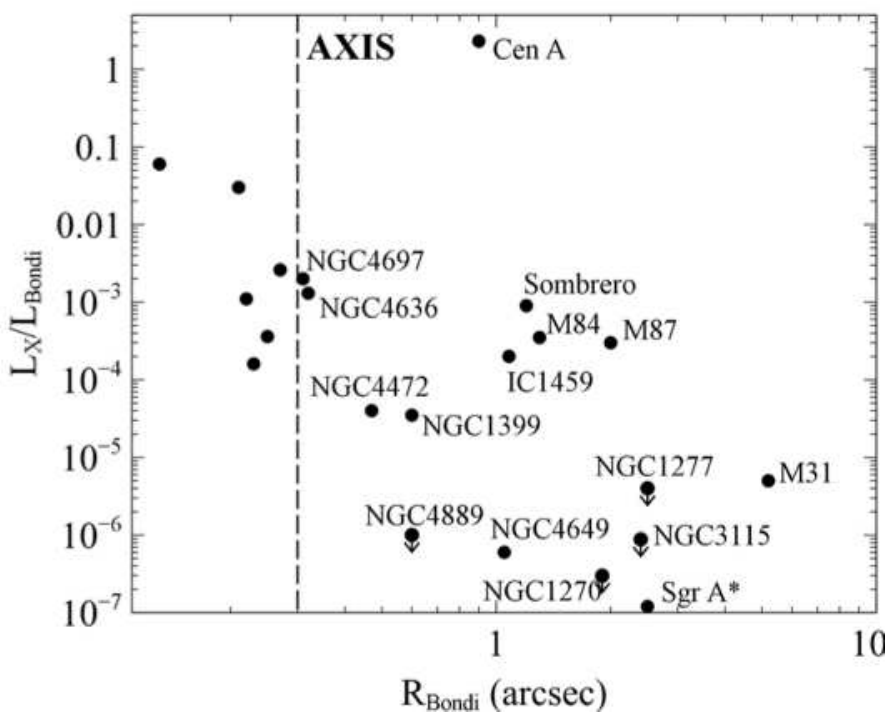}
   \end{tabular}
   \end{center}
   \caption[example] 
   { \label{bondi_radius} 
The host galaxies for which AXIS, with 0.3$''$ resolution, can measure the surface brightness and temperature distribution for emission inside the Bondi radius.}
   \end{figure} 

Resolving the structure and dynamics of the inflowing gas, once it falls within the gravitational influence of the black hole, is  essential for understanding the accretion flow which is bounded by the Bondi radius R$_{B}$ = 2GM$_{BH}$/c$_{s}^{2}$  or R$_{B}$ $\sim 0.2$T$_{keV}^{-1}$M$_{9}$D$_{10}$ arcsec, where M$_{9}$ is the mass of the black hole in units of 10$^{9}$ $\msun$, D$_{10}$ is the distance of units of 10 Mpc, and T$_{keV}$  is the temperature of the gas near R$_{B}$.  In the absence of angular momentum, the black hole is predicted to gravitationally capture the ambient ISM surrounding it at a rate $\dot{M}_{Bondi}$ = 4$\pi$$\lambda$(GM$_{BH}$)$^{2}$nc$_{s}^{-3}$, where $\lambda =0.25$ for an adiabatic process and $n$ is the density of the gas at R$_{B}$.

The Bondi region can be resolved by Chandra in only a handful of the nearest objects.  Sgr A*, NGC3115 and M87 were targeted in a series of Large and Visionary Chandra projects \cite{Wang13, Wong14, Russell18}.  These observations revealed shallow gas density profiles inside the Bondi region and imply the presence of outflows that expel the vast majority of the matter initially captured by the black hole.  In M87, Chandra also uncovered a rapid transition in the hot gas structure within the Bondi radius where the radiative cooling time of the gas decreases sharply to only 0.1$-$0.5 Myr, which is comparable to the free fall time.  In strong contradiction to most theoretical calculations, these observations suggest that the gas flows within the Bondi radius are a complex mixture of rapidly cooling inflow fueling the black hole and powerful jet-driven outflows.  However, Chandra's angular resolution and limited collecting area at low energies are barely adequate to this task.

With an effective angular resolution of 0.3$''$ and increased sensitivity, there are at least 25 supermassive black holes in the local universe \cite{Garcia10, Vandenbosch16} for which the Bondi radius can be resolved (see Figure~\ref{bondi_radius}).  AXIS will map the detailed density and multi-temperature structure within the black hole's sphere of influence to reveal transitions in the inflow that ultimately fuel the black hole activity, outflows along the jet-axis that limit the accretion rate and thereby build up the first detailed  picture of these accretion flows.

\subsubsection{Dual AGN}

The general theory of structure formation predicts that mergers are a major component of galaxy growth and evolution.  Since almost all massive galaxies at low redshift contain central supermassive black holes, it has long been predicted that when the galaxies merge, so should their black holes. However, this prediction has been very hard to verify since the timescales for the BH merger to occur are very uncertain. One of the few observational tests of this idea other than gravitational waves is to search for "dual black holes" in nearby galaxies \cite{Steinborn16}. Theoretical calculations indicate that a significant fraction of these sources would be active galaxies, which has stimulated an intensive search for dual AGN \cite{Koss12}, but despite intensive work there is little information about the occurrence rate of dual AGN for a large sample of objects covering a wide range in mass, luminosity and nature of the host galaxy critical for constraining model of gravitational wave sources. 
  
Dual AGN are extremely rare in the radio \cite{Burkespolaor11} and, optical selection techniques for dual AGN are extremely inefficient, producing a very large fraction of "falses" \cite{Nevin16}. In contrast, Chandra X-ray observations have discovered all three of the "dual AGN" with the closest separation and have found a significant fraction of all the believable candidates \cite{Koss15}.  However, the limited Chandra resolution and sensitivity have severely restricted the redshift range in which duals can be found and the range of relative intensities in which they can be detected.

   \begin{figure}[htb]
   \begin{center}
   \begin{tabular}{c}
   \includegraphics[height=8cm]{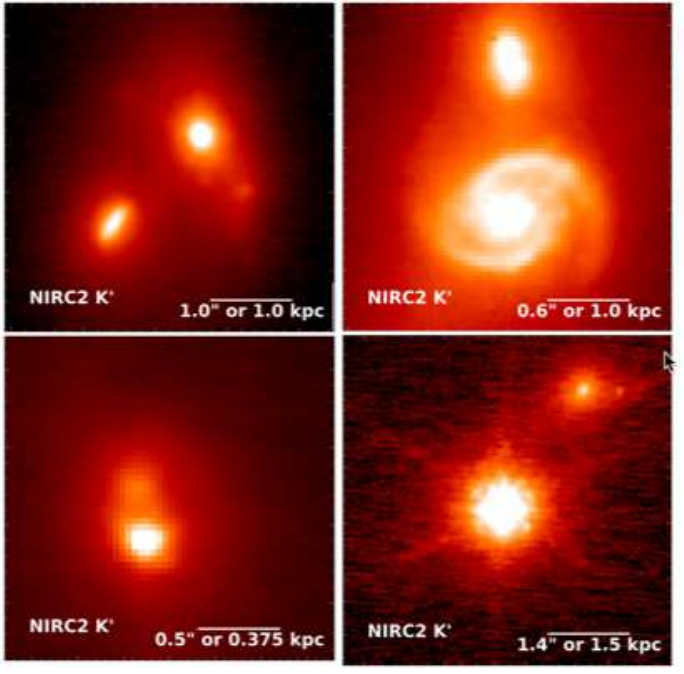}
   \end{tabular}
   \end{center}
   \caption[example] 
   { \label{nuclear_mergers} 
Adaptive optics IR images of nuclear mergers (Koss et al. 2018, submitted)}
   \end{figure} 

Recent high resolution NIR observations of the host galaxies of luminous obscured AGN have found that very close mergers are common (Figure~\ref{nuclear_mergers}), but IR data do not reveal which of these nuclei are AGN.  AXIS observations would answer the critical questions: (i.) is the low observed rate of dual AGN consistent with the theoretical merger predictions upon which LISA and pulsar timing arrays (PTA) are based; (ii.) what is the frequency, environment, and luminosity dependence of dual AGN; (iii.) is the obscuration level of the AGN correlated with merger stage; (iv.) How are mergers related to the luminosities of the sources; and (v.) how does accretion rate correlate with merger stage and the dual AGN rate? The kpc-scale dual AGN population that will be well sampled by AXIS provides key constraints for the SMBH mass merger function for the expected LISA and PTA SMBH merger rates \cite{Kelley18} 

\subsubsection{The Highest Redshift Universe}

The very high-redshift universe has largely been beyond the sensitivity limits of Chandra and XMM-Newton with only a small number of X-ray AGNs have been directly detected beyond z$>$5.  In combination with WFIRST data, AXIS has the potential to generate much larger samples of z$>$5 AGNs, allowing the study of how the AGN luminosity function changes at high redshifts and comparison with the star formation history, which may drop off more slowly than the AGN luminosity function.  This is a key measurement in understanding the sequencing of the growth of supermassive black holes relative to the growth of galaxies. 

At even higher redshifts, z=6$-$8 there are only upper limits of $\sim 2$ $\times$ 10$^{41}$ erg s$^{-1}$ on the average X-ray luminosity of optically selected galaxies, below what is expected for rapidly growing black holes. Since this is an average, if there are also significant AGN contributions,  as predicted, AXIS should be able to detect individual sources, constraining the growth of black holes at the highest redshifts.  AXIS will be able to measure the emission from virtually all the JWST very high-z SMBH candidates and thus either confirm them or show that they are imposters. 

These observations are key to understanding the early growth of supermassive black holes and what typical masses they have grown to at these redshifts.  The AXIS angular resolution will allow the separation of AGNs from the more extended star-forming contributions and  permit unique optical/IR counterparts to be determined.

\subsection{Galaxy Clusters and Cosmic Large Scale Structure}

   \begin{figure}[htb]
   \begin{center}
   \begin{tabular}{c}
   \includegraphics[height=7cm]{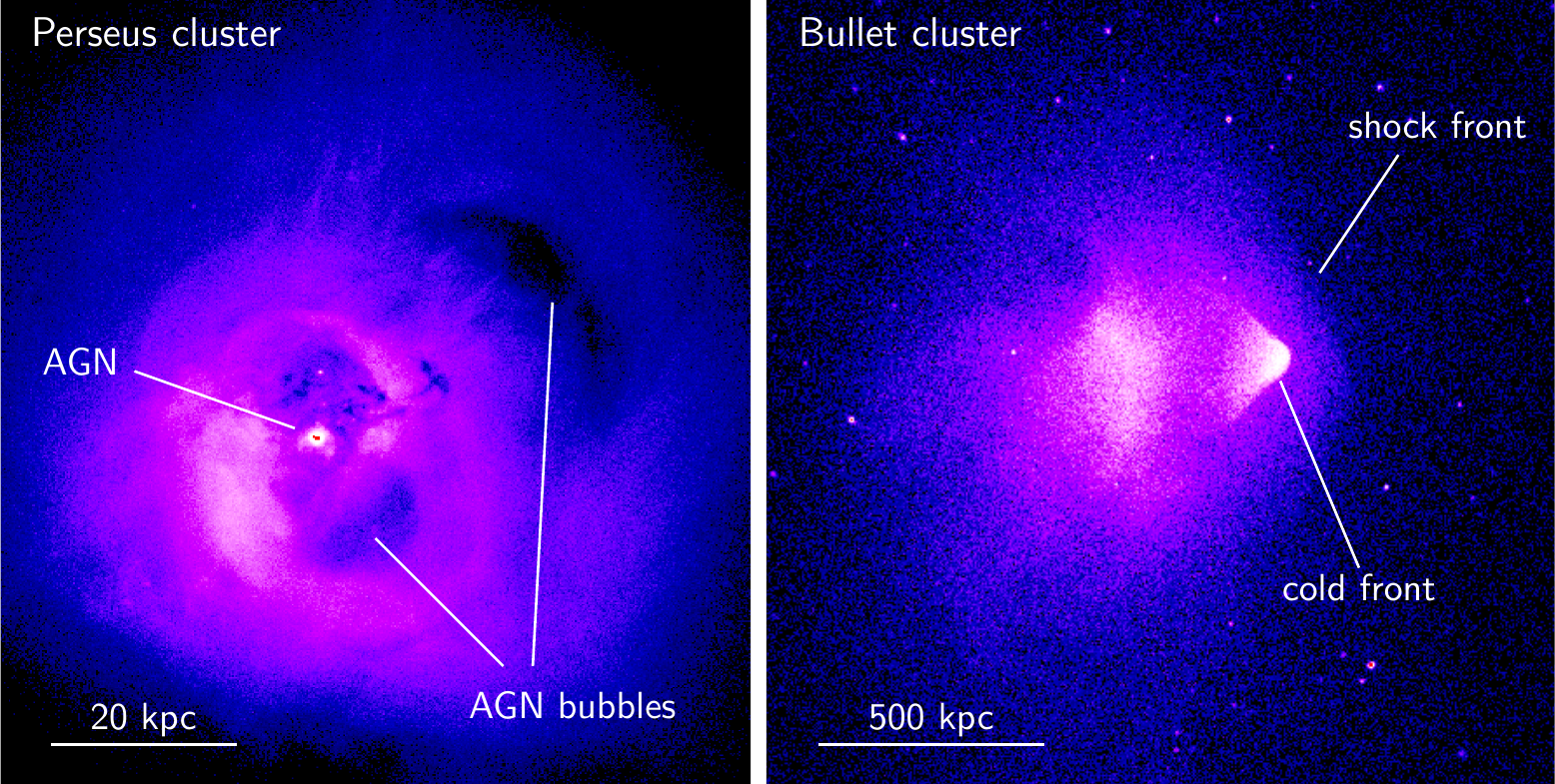}
   \end{tabular}
   \end{center}
   \caption[example] 
   { \label{chandra_clusters} 
Chandra high-resolution X-ray images of galaxy clusters. (a) Cool core of the Perseus cluster, harboring a supermassive black hole that ejects relativistic matter into the cluster plasma. Rich details of interaction of the jets with thermal gas can be seen. (b) Bullet cluster, experiencing a violent merger. The cool bullet ("cold front") drives a shock front, which converts vast kinetic energy of the merger into thermal energy and feeds magnetic fields and cosmic rays. High-resolution X-ray observations of these phenomena enable unique tests of plasma physics.}
   \end{figure} 

Galaxy clusters provide us with a unique set of tools to study cosmology, plasma physics, cosmic particle acceleration, galaxy formation, and feedback from the accreting supermassive black holes that lie in the cluster centers. Clusters are the biggest gravitationally bound, dark-matter dominated objects in the Universe, whose number density is exponentially sensitive to the cosmological model and whose matter content is representative of that of the Universe as a whole. They are filled with hot, magnetized plasma -- their dominant baryonic component -- that is optically thin in X-rays, so their X-ray images and spectra yield robust constraints on physical models.

   \begin{figure}[htb]
   \begin{center}
   \begin{tabular}{c}
   \includegraphics[height=8cm]{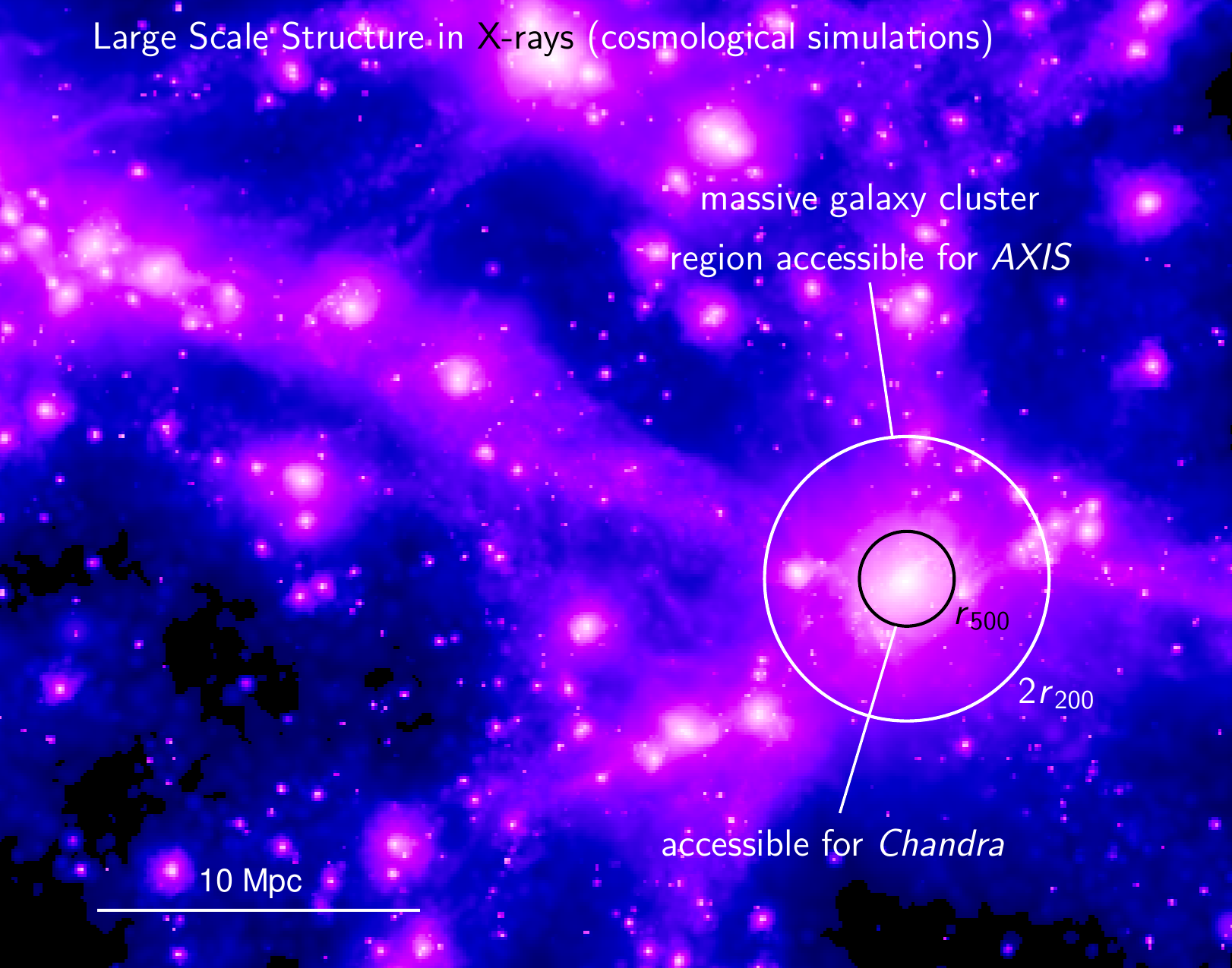}
   \end{tabular}
   \end{center}
   \caption[example] 
   { \label{largescale_simulation} 
Cosmological simulations of Large Scale Structure \cite{Dolag06}. Color shows X-ray brightness of the intergalactic plasma, showing a web of giant filaments with galaxy clusters as bright nodes. Chandra and XMM are limited to bright, mostly hydrostatic regions such as the one within the r$_{500}$ circle. AXIS will be able to go to much lower brightness and study the dynamic regions at the interface between clusters and the Cosmic Web.}
   \end{figure} 

Chandra high-resolution imaging has uncovered a wealth of physical phenomena in clusters (Figure~\ref{chandra_clusters}), from mechanical interaction of powerful AGN outbursts with the intracluster plasma in cluster cores \cite{Fabian06, Forman07} to shock fronts and "cold fronts" that occur during violent cluster collisions --- the most energetic events in the Universe since the Big Bang \cite{Markevitch07}. However, these studies have been limited to the cluster inner, brighter regions, and even there, some important physical measurements remain out of reach. AXIS will make dramatic contributions in many areas of cluster astrophysics; just a few examples are given below.

\subsubsection{Feedback from Supermassive Black Holes}

In the cores of nearby clusters, Chandra has detected AGN-blown X-ray "cavities" (seen as bright lobes in the radio), through which the supermassive black hole is thought to regulate accretion onto itself --- so-called AGN feedback (Figure~\ref{chandra_clusters}). However, there are few constraints on the composition of the relativistic matter ejected from the AGN that displaces the thermal plasma. AXIS will have the resolution and the collecting area to observe such cavities at redshift z $\sim 1$, where a cavity turns into an X-ray bright spot, because the surface brightness of inverse Compton emission increases with the CMB energy density as (1+z)$^{4}$. These data will unambiguously determine the density of relativistic electrons and magnetic fields in the AGN bubbles, and thus determine the critical physical parameter, the energy budget of the AGN outburst. Over the cluster lifetime, the AGN bubbles float buoyantly from the cluster core to the outer regions, expand, mix and become almost invisible. AXIS will have the sensitivity to detect those very low-contrast, giant "ghost cavities," in collaboration with radio observatories such as LOFAR and SKA. This would provide a full account of the energy output over billions of years of a supermassive black hole's growth and thus strongly constrain all feedback models.

\subsubsection{Shock Fronts, Particle Acceleration, and Plasma Physics}

The study of cluster shock fronts --- surfaces where energy is exchanged between the gas flows, thermal plasma, magnetic fields and cosmic ray particles --- has a major impact on our basic understanding of plasma astrophysics, providing one of the best natural laboratories. AXIS will provide the low detector background, the angular resolution to resolve away the discrete CXB, and the high photon statistics, a combination required to (a) find shock fronts far out in the cluster outskirts, where most of them should reside, and (b) study the fine structure of those shocks. Figure~\ref{largescale_simulation} compares the reach of AXIS to that of Chandra for detailed studies of the various cluster phenomena, including shocks. Clearly, AXIS will uncover much more of the  structure-formation action in clusters than currently achievable.  Resolving bow shocks such as the one in the Bullet Cluster (Figure~\ref{chandra_clusters}) but at higher redshifts, co-located with synchrotron "radio relics" produced by the shock-accelerated ultrarelativistic electrons in the shock-amplified magnetic field, would provide detailed data on the interactions between thermal and relativistic plasma components. Synergy between AXIS on the X-ray side and LOFAR and SKA on the radio side would make such studies possible. AXIS will be able to detect a temperature precursor to the shock, and possibly the cosmic-ray pressure precursor for the first time, and determine thermal conduction and cosmic ray diffusion in the magnetized intracluster plasma. AXIS will also measure the electron-proton equilibration timescale \cite{Markevitch07} for a large sample of shock fronts, eliminating the inherent systematic geometric uncertainty. Angular resolution and photon statistics are critical here as well. Collisionless shock fronts in the hot magnetized cluster plasma probe a different regime from those in the solar wind and other astrophysical systems. The above measurements will have importance not only for cluster astrophysics but for plasma physics in general.

\subsubsection{How Galaxy Clusters Connect to the Cosmic Web}

   \begin{figure}[htb]
   \begin{center}
   \begin{tabular}{c}
   \includegraphics[height=8cm]{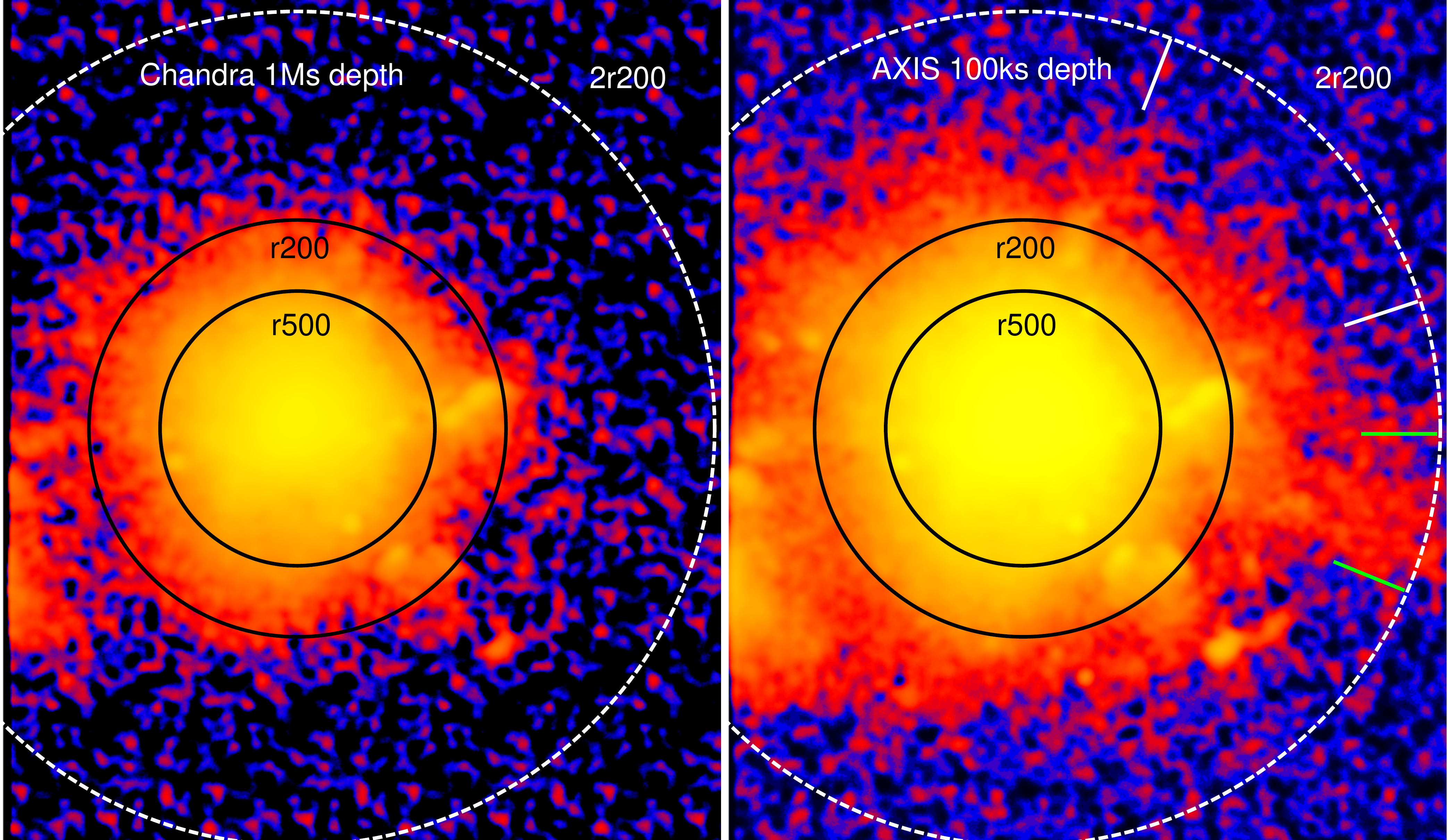}
   \end{tabular}
   \end{center}
   \caption[example] 
   { \label{massive_cluster_simulation} 
A massive cluster from cosmological simulations \cite{Dolag06} as observed by Chandra and AXIS. Exposures are selected to yield the same number of cluster counts. Because of much lower background, AXIS will detect the X-ray emission much farther from the center --- where clusters connect to the Cosmic Web.}
   \end{figure} 

The intergalactic medium, IGM (a.k.a. WHIM, warm-hot intergalactic medium) is one of the fundamentally important, yet unexplored areas in astrophysics. The IGM is believed to be the main reservoir of the "missing baryons" in the local Universe, and the ultimate depository of metals and entropy produced in galaxies over the cosmological time. Observations of IGM have long been attempted, but so far, only a tiny fraction of it --- the relatively dense, cold phase visible through the OVI absorption in FUV --- has been unambiguously detected. AXIS will open a large discovery space here. Galaxy clusters act as "IGM traps," attracting and compressing the IGM and making it potentially observable in X-ray emission. However, its surface brightness is extremely low, requiring an instrument with a very low background such as AXIS to measure it. Removal of the discrete CXB is critically important, as its brightness is 1.5$-$3 orders of magnitude above the cluster brightness at radii (1-2)r$_{200}$. With these properties AXIS will trace the IGM in regions far beyond the cluster virial radius (r$_{200}$) where the IGM flows into clusters along the giant filaments of the Cosmic Web (Figure~\ref{largescale_simulation}). Therefore, we find the infalling galaxy-sized and group-sized clumps (with the accompanying shock fronts and gas stripping), as well as the IGM filaments as hinted at by ultra-deep Chandra observations.

\begin{figure}[htb]
   \begin{center}
   \begin{tabular}{c}
   \includegraphics[height=7cm]{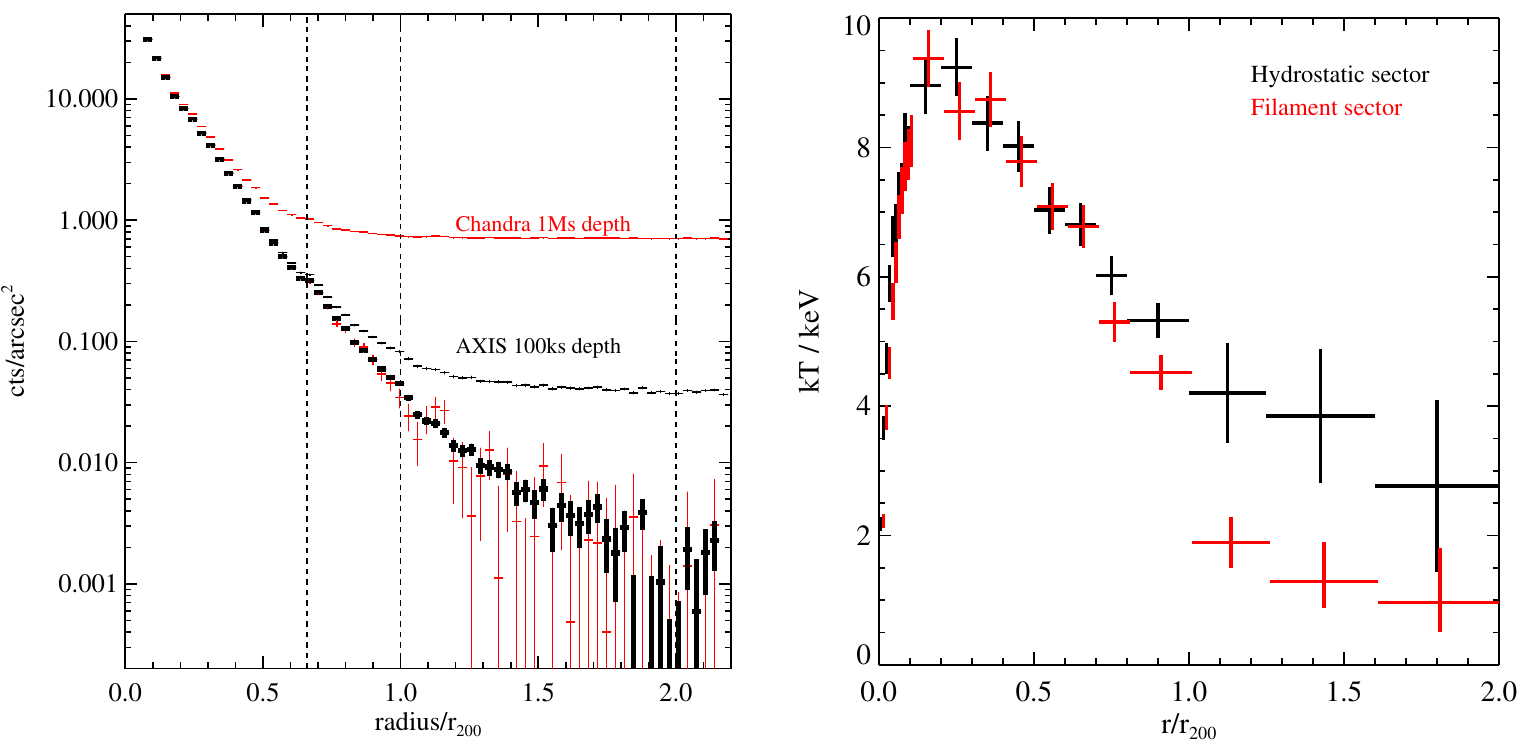}
   \end{tabular}
   \end{center}
   \caption[example] 
   { \label{radial_profiles_simulation} 
X-ray surface brightness and temperature profiles for the simulated
galaxy cluster shown in Figure~\ref{massive_cluster_simulation}. {\it Left:} Surface brightness profiles (0.8$-$5 keV) extracted in a sector away from the filament, shown in Figure~\ref{massive_cluster_simulation} with white dashes. Profiles before and after subtracting the detector background and the soft diffuse CXB are shown (both instruments equally resolve away the discrete CXB). Dashed lines are at r$_{500}$, r$_{200}$ and 2$r_{200}$. AXIS can trace the emission well into the region where the cluster connects into the Cosmic Web. {\it Right:} Temperature profiles for two sectors shown in Figure~\ref{massive_cluster_simulation} with white and green dashes. A background uncertainty of $\pm$15\% is included. Temperatures in the filaments will be measured with high accuracy.}
   \end{figure}

In Figure~\ref{massive_cluster_simulation}, we compare the simulated X-ray brightness of a massive nearby galaxy cluster for AXIS and Chandra, for exposures that have the same number of cluster photons in the 0.8-5 keV energy band, where the Galactic background is low. Radial X-ray brightness profiles in a "hydrostatic" cluster sector unaffected by the IGM filament are shown in Figure~\ref{radial_profiles_simulation}. The Chandra signal disappears in the irreducible detector background right at the virial radius, while AXIS will trace the emission at least twice as far. For the first time we will see directly how the galaxy clusters connect to the diffuse Cosmic Large-Scale Structure, leveraging AXIS's spectacular surface brightness sensitivity.

Not only will we trace the X-ray brightness but measure the plasma temperature as well, particularly accurately for the bright regions of IGM filaments (Figure~\ref{radial_profiles_simulation}). For the "hydrostatic" cluster regions, the temperatures will provide an estimate of the total cluster mass enclosed within that radius. Comparing it with other mass estimates, such as those from gravitational lensing from Euclid and WFIRST, will allow estimates of the nonthermal pressure components in the plasma. We will also be able to detect the 1 keV Fe line in IGM filaments near clusters, serving as a pathfinder for the Athena and Lynx calorimeters that will have much smaller FOV. By uncovering the prevailing conditions in the IGM, these studies will directly confront the galaxy formation and feedback models.

\subsection{Time Domain Science}

As evidenced by the 2010 Astrophysics Decadal Survey ranking of the Large Synoptic Survey Telescope as the highest priority project for ground-based astronomy, time-domain astronomy is an essential component in understanding our universe. When AXIS launches in 2028, there will be several new observatories, across different EM wavebands and different messengers (e.g. low and high frequency gravitational waves, neutrinos and cosmic rays), which will routinely find large samples of time variable extreme astrophysical phenomena out to higher redshifts than currently possible. AXIS will have a Target of Opportunity response time of less than 4 hours to follow-up transients discovered by other facilities and will have unprecedented X-ray sensitivity (down to 10$^{-18}$ erg cm$^{-2}$ s$^{-1}$; 10$^{4}$ more sensitive than Swift XRT), providing the much-needed X-ray constraints that can identify and characterize new transients. 

Tidal disruption events (TDEs) were first discovered in soft X-rays with the ROSAT All-Sky Monitor in the mid-1990s \cite{Komossa15}, and were characterized with a thermal spectrum at 10$^{5}$ K, which is thought to be emission produced from a newly formed high-Eddington or even super-Eddington accretion disc. It is not at all clear what the origin of the optical emission, whether from reprocessed disc emission or from shocks caused by the self-intersecting debris stream. X-ray follow-up of optically-selected TDEs is essential for understanding the origin of the TDE emission. Currently only about 2 dozen TDEs have been found, but by the late 2020s, LSST will be online and predicts to see $\sim 1000$ TDEs per year. Furthermore, the Square Kilometer Array should detect the subset of TDEs that produce relativistic jets ($\sim 500$ predicted  yr$^{-1}$). TDEs will be detected to much larger redshift, and therefore it is essential to have a high sensitivity X-ray instrument that can confirm the location of the X-ray emission with the nucleus of the galaxy and measure its spectrum and time behavior. We have very few constraints on when the X-ray is produced, relative to the optical detection, and therefore rapid sensitive X-ray follow-up is required. 

The recent discovery of gravitational waves from a binary neutron star merger, and the subsequent EM detections in the form of a short Gamma-Ray Burst and kilonova emission, has opened a new field of observational astronomy. X-ray follow-up to binary neutron star mergers can put constraints on the binary inclination, which, in the gravitational wave signature, is highly degenerate with the distance of the object. Based on results from GW170817, one requires a highly sensitive, fast-response X-ray mission to follow-up significantly more distant gravitational wave events from binary neutron star mergers. High angular resolution is required in order to distinguish between the host galaxy nucleus and the merger event as shown in the x-ray images of GW170817.

Finally, the LISA space-based gravitational wave detector is planned to launch in the early 2030s, during the planned mission time of the AXIS mission, and with it, to measure gravitational waves from the in-spiral and coalescence of supermassive black holes binaries (SMBHBs). If one or more of the black holes is actively accreting, one expects strong X-ray signatures. LISA will be sensitive to gravitational wave signatures of SMBHBs at very high redshift for several orbits before final coalescence, which corresponds to timescales of months to years, thus requiring high sensitivity and high angular resolution in order to pinpoint the emission to the galaxy?s center. The rapid slew capabilities of AXIS will allow for long term monitoring of SMBHB systems, until the point of final coalescence.

AXIS will have the fast slew rate and ToO response time, high sensitivity and high angular resolution required to obtain these science goals, and will have a significant percentage of the primary mission dedicated to rapid follow-up and study  of transients.  

\subsection{Chemical Evolution of the Universe}

\subsubsection{Supernova Remnants and Supernovae}

The remnants of supernovae are amongst the most complex X-ray sources, ranging in extent from arcseconds to degrees in diameter, with shock speeds from tens to tens of thousands of km s$^{-1}$ and plasma conditions covering a wide range of possibilities. Chandra, XMM-Newton, and Suzaku have allowed great progress in this field, yet major scientific questions still remain: such as 1) How do supernovae dictate the life cycle of elements in the ISM? 2) What are the progenitors of the various types of supernovae (Type Ia, Type Ib/c, Type II, etc.), and how do they relate to the remnants produced? 3) How are particles accelerated to extraordinary energies in shock waves? Answering these questions require a high-spatial resolution high sensitivity X-ray imager, to be combined with the high-spectral resolution instruments of the next decade and beyond. Here, we point out specific science goals that only a high-spatial resolution mission like AXIS could achieve.

{\bf Supernova Ejecta Mapping}

The wide range of explosion models for both Type Ia and core-collapse (CC) SNe predict vastly different ejecta distributions in the expanding remnant, but measuring this distribution requires both high angular resolution and good sensitivity. Chandra has made such ejecta line maps for a few remnants (for example, see \cite{Hwang00} for maps of Cas A, \cite{Warren05} for Tycho's SNR, \cite{Park02} for G292.0$+$1.8, or \cite{Winkler14} for SN 1006).  However, such mapping, particularly in smaller remnants such as those in the LMC, as well as in the faint Fe K$\alpha$ band, is difficult for Chandra to obtain but is an important discriminator between various supernova explosion models.

   \begin{figure}[htb]
   \begin{center}
   \begin{tabular}{c}
   \includegraphics[height=8cm]{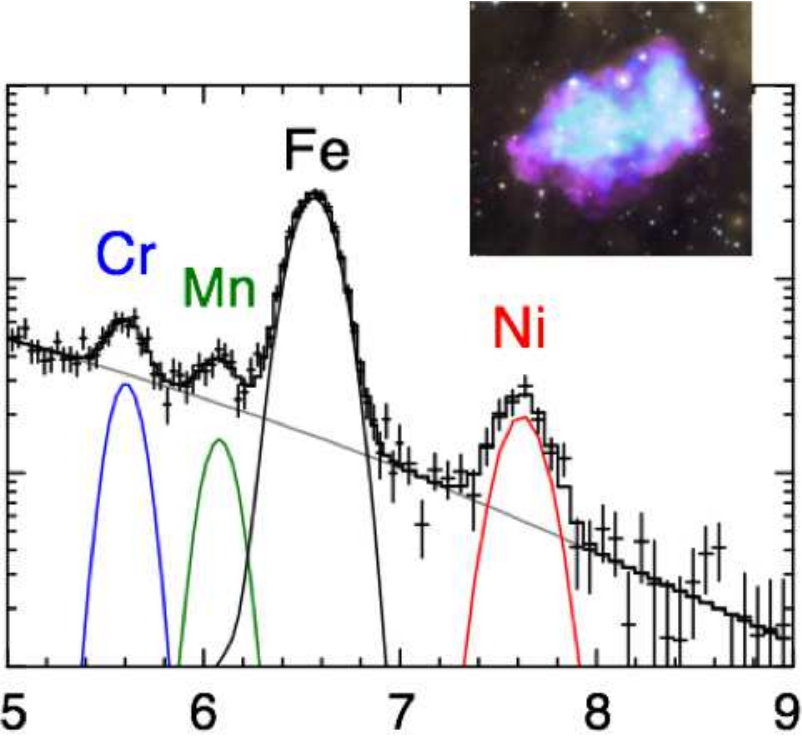}
   \end{tabular}
   \end{center}
   \caption[example] 
   { \label{fe_group_3c397} 
Suzaku spectrum of the Type Ia Galactic SNR 3C397 \cite{Yamaguchi15}, showing the Fe-group elements, with a Chandra image of the remnant inset. The elevated Cr, Mn, and Ni abundances, with respect to Fe, imply an explosion in a Chandrasekhar-mass white dwarf, hinting at an origin in a single-degenerate system.}
   \end{figure} 

Critically important \cite{Yamaguchi15} is mapping in the Fe K band where lines of the Fe-group elements of Cr, Mn, and Ni are located. This allows determination of the mass of the Type Ia progenitor, a critical measurement for galactic metallicity evolution models. These lines are quite faint and difficult to observe with Chandra or XMM-Newton. The best spectrum obtained (Figure~\ref{fe_group_3c397}) is the integrated spectrum of the entire remnant, meaning that while the total abundance of these elements can be measured for a few bright objects, their spatial distribution, another key discriminant in distinguishing various explosion models \cite{Seitenzahl13}, cannot be mapped by present instrumentation.

{\bf Shock Wave Physics and Particle Acceleration}

Shock waves are ubiquitous in the universe, and SNRs offer the chance to observe shock fronts on small (sub-parsec) scales with a range of Mach numbers. This offers several avenues for further study beyond what Chandra has enabled. For instance, shocks are almost certainly responsible for producing high-energy cosmic-rays through turbulent scattering off of amplified magnetic fields, but the detailed process for this is still under study. Virtually all theories predict that in shocks dominated by non-thermal synchrotron emission, some faint X-ray emission should be present ahead of the shock from particles that have diffused upstream, yet this emission has never been detected due to the limited capabilities of present day observatories. Finding and characterizing this precursor in   Galactic remnants with synchrotron-dominated shock waves will be a key goal for a high-sensitivity, low-background mission like AXIS, and will put tight constraints on the universal properties of fast shock waves, such as the degree of magnetic field amplification, and the diffusion and scattering length of energetic particles.

{\bf Proper Motion Observations}

By observing the time evolution of SNRs, by measuring  their expansion through proper motions AXIS will build on Chandra's legacy, not only opening up new remnants for expansion measurements, but refining previous measurements as the uncertainties will decrease with increased time baselines. Such observations will determine the three-dimensional velocities of dozens of tufts of ejecta,  allowing comparisons to 3D models of SNe explosions, as was done in Tycho's SNR \cite{Williams17}. Chandra data on Tycho showed that it was the result of a largely symmetric explosion, favoring multiple ignition points within the progenitor.  AXIS observations will allow applications of this technique to other remnants and other ejecta species not possible with Chandra. With more than an order of magnitude increase in X-ray count rates and lower background levels, AXIS would perform these measurements on many remnants in the Galaxy, of both Type Ia and CC origin.

{\bf SNe and SNRs in Nearby Galaxies}

The large effective area of AXIS and high angular resolution will enable systematic studies of SNRs in nearby galaxies, such as M31 and M51. As an example of what is possible, \cite{Long14} used 730 ks of Chandra observations to observe M83 and found, 87 SNRs (compared to 225 optically-detected SNRs in M83). However, the signal to noise was not sufficient for spectroscopy. AXIS would make this possible, and would thus allow for a comparison of the SNR population of many nearby galaxies with the local star-formation history. 

Follow-up of SNe in galaxies within several Mpc in the past decade or so, well before they reach the remnant phase, is extremely important for SN progenitor models and requires high spatial resolution and sensitivity.  X-rays are produced in core-collapse SNe due to shock interaction with the dense circumstellar medium surrounding the explosion and can persist for several years. Monitoring of the light curves from these young SNe places strong constraints on the amount of mass lost by the progenitor system, a key parameter of SNe models.  

Finally, an exciting potential exists if we can catch a core-collapse SN within the first few hundreds of seconds post-explosion, potentially possible given the rapidly slewing spacecraft. Core-collapse SNe can reach brightnesses of 10$^{44}$ ergs s$^{-1}$ at timescales that range from seconds after the explosion for Wolf-Rayet stars to thousands of seconds for red supergiants \cite{Nakar10}. Catching the immediate afterglow of a the shock breakout in a core-collapse SN has only been done a handful of times, perhaps most notably in SN 2008D \cite{Soderberg08}. Combined with a quick slew capability, AXIS will achieve enough sensitivity to detect SNe out to several hundred Mpc.

\subsubsection{Dust Scattering Halos}

AXIS will detect dust halos, that is, X-rays from a bright source that have been scattered into a halo by intervening dust grains. The halo brightness depends on the large end of the grain size distribution, grain composition, and location along the sightline \cite{Mathis91}; the closer that grains are to the source, the smaller the radius of the resulting halo. AXIS's exquisitely high resolution and sensitivity measurement much closer to the X-ray source and at higher distances than ever before, including studies of time delays in halos formed around bright AGN by an intervening dusty galaxy \cite{Corrales15}. These time delays can be used to find the distance to the galaxy, thus allowing the calibration of stellar luminosities and measurement of the Hubble constant, H$_{0}$ \cite{Draine04}. This new method offers an independent check, relevant in light of the current tension in the derived values of H$_{0}$ from more traditional methods. With AXIS, the distance to M31 can be determined  to 1\% accuracy in dedicated observations of 1 Msec.
					
AXIS will perform detailed  studies of dust halos in our own galaxy, the LMC and the SMC, providing badly needed constraints on the large grain population and compositions in these regions, and thus on models of interstellar dust \cite{Mathis77, Predehl95, Zubko04, Jenkins09, Valencic15}. Knowledge of Galactic grain properties is particularly important, given the impact of dust on measurements of the CMB polarization and the fact that the expected polarized dust emission in the microwave regime is highly dependent on grain size and composition \cite{Draine09, Draine13}.  Measuring the population of large grains in the local ISM is also important for understanding the interplay between destructive (shocks and sputtering) and constructive (metal enhanced outflow) processes in the Universe, providing a more complete picture on the origins and life cycle of interstellar dust \cite{Jones11, Dwek16}.

With its supreme soft X-ray sensitivity and high resolution capabilities, AXIS will be able to image the full 10-15$'$ extent of Galactic dust scattering in high contrast, allowing the identification of individual clouds along the sight line. Currently, this type of X-ray tomography has only been achieved by examining dust echoes from the brightest of X-ray binary outbursts \cite{Heinz15}. CCD spectra will obtain an SED of dust scattering halo from individual clouds, which directly measures the upper end of the grain size distribution \cite{Corrales15}. For X-ray binaries with known distances, dust scattering echoes can be used to map the line of sight dust distribution with greater precision than current stellar population extinction maps \cite{Heinz16}.

\subsection{Solar System Science}

X-ray emissions has been detected from a variety of different solar system objects, including, planets, satellites, and comets \cite{Bhardwaj07}. These emissions are primarily generated from interactions between the host object and solar photons, solar wind outflows, cosmic X-ray background, or a combination of these sources \cite{Cravens00, Bhardwaj07} providing insight into both the observed solar system object and the surrounding X-ray medium.

AXIS is uniquely poised for solar system science in a manner not possible with previous X-ray observatories. Earlier X-ray observations have been hindered by poor count statistics, necessitating time averaged results from deep exposures \cite{Lisse96, Lisse01, Dennerl97, Dennerl02, Wolk09, Snios16}. The large effective area of AXIS will allow for investigations of temporal variations in X-ray emission on time-scales comparable to target rotation rates, orbital periods, and solar X-ray fluctuations \cite{Neupert06}. This will be the first time a thorough time-domain survey of solar system objects has been performed in X-rays. Furthermore, the high spatial resolution will provide high-quality images that may be used to generate chemical composition maps of neighboring bodies. For example, AXIS will be able to map the abundances of O, Mg, Al, Si, and Fe on the Moon at a 2 km surface resolution using only a 100 ksec exposure, which is an increase of 10 in spatial resolution over the current best surface resolution \cite{Yokota09}. Utilizing both the spatial and time resolution of AXIS will allow us to probe both the transportation of elements and chemical reactions in a system due to high-energy activity, like solar flare impacts on planetary upper atmospheres. Such maps will provide crucial knowledge into the short- and long-term evolution of planetary atmospheric composition and asteroid resource exploitation, exciting new capabilities in the current era of planetary research \cite{Glassgold12, Cleeves17}. 

\subsection{Complementarity and Expansion of Phase Space}

AXIS would vastly increase sample sizes by factors of 10$-$30 over Chandra allowing large samples of almost every class of celestial object.  As is very common in astrophysics is it only by studying physical phenomena with a reasonable sized sample that strong conclusions can be reached.  With the $>$ 0.5 Ms Chandra exposures required for detailed studies of almost every extended source, the Chandra samples are small and highly biased to the highest surface brightest members of the population. 

AXIS will have complementary angular resolution and sensitivity to next generation observatories in other bands. All of the next generation observatories (ALMA, WFIRST, JWST, SKA, adaptive optics on 30m telescopes) have $<$1$''$ angular resolution. In addition, many of these facilities have vastly improved sensitivity compared with the previous generation of telescopes for which Chandra was a good match.  Without a high angular x-ray mission with adequate capabilities operating at the same time as these new observatories, the rate of major advances will drastically diminish from the number we have made over the last two decades.

\subsection{Other Important Science}

There are many other science topics that could be discussed but for which we did not have space in this paper.  We hope that the community will join in the discussion during our public workshop in August 2018.  Amongst these are detailed study of star clusters, resolving supernova remnants in nearby galaxies, the astrophysics of jets, detailed maps of starburst galaxies, deep X-ray surveys, pulsar wind nebulae, star formation regions in the local group, cluster cooling fronts, galaxy interactions, bubbles in clusters, ram pressure stripping of galaxies in groups and clusters, particle acceleration in SNR, cluster structure at high z and comparison with the S-Z effect, observations of planets, interaction of AGN with cluster gas at moderate to high z, identification and timing of X-ray binary hosts within 5 Mpc, and expansion of SNR in the LMC/SMC.   Those topics discussed here capture the driving technical requirements for AXIS.  

\section{Technical Section}

\subsection{Instrumentation}

The AXIS instrumentation consists of a high resolution X-ray mirror assembly and a focal plane CCD or similar type detector.  As described below, the primary technology development driver is the mirror, while the detector requirements are similar but considerably improved  to the capabilities of present day devices. 

\subsubsection{Mirror}

The X-ray mirror assembly for AXIS combines large throughput with high angular resolution. The single crystal silicon mirror technology currently under development at Goddard Space Flight Center \cite{Zhang18}  is expected to meet the four-fold requirement of PSF, effective area, mass, and production cost. 

The AXIS mirror has a 9 m focal length, a 1.8 m outer-diameter, and a total 298 shells.  The axial length of each shell is 100 mm for the primary and 100 mm for the secondary. Each shell is segmented into many segments in the azimuthal direction such that each mirror segment is approximately 100 mm by 100 mm, resulting in 16,538 mirror segments. The 298 shells are grouped into 6 meta-shells, each of which is composed of a structural shell and approximately 2,700 mirror segments which are precisely aligned and bonded together with and epoxy and silicon spacers, resulting in a total mass of approximately 450 kg. The mirror segments are fabricated out of mono-crystalline silicon using state-of-the-art polishing techniques and are coated with 30 nm of iridium to enhance their X-ray reflectance, providing more than 7,700 cm$^{2}$ of effective area at 1 keV and 1,600 cm$^{2}$ at 6 keV. The short axial length of the mirror segments and mounting the mirror pairs conforming to the principal surface of a Wolter-Schwarzschild design optimizes the off-axis imaging performance, resulting in an order of magnitude improvement in off-axis PSF compared with Chandra.

The mirror technology development at GSFC was initiated by the Constellation-X project in the early 2000s was continued by the subsequent International X-ray Observatory project until 2011 and has continued to be funded through NASA's Strategic Astrohpysics Technology program to enable future missions such AXIS and Lynx.  The use of mono-crystalline silicon as the mirror substrate material enables the use of direct fabrication of thin ($<$1 mm) and lightweight (areal density $<$ 2 kg m$^{-2}$) mirrors with the highest possible angular resolution.  The primary technical challenge lies in perfecting a process to minimize production cost and weight by more than a factor of 50 on a per unit area basis with respect to Chandra's mirror.  Individual mirror segments with 0.5$''$ PSF have been fabricated, comparable to Chandra's. It is likely that individual mirrors segments meeting AXIS requirements will be made by 2020.  A mirror alignment and bonding process is being developed in parallel.The basic elements of this process have been demonstrated in early 2018, resulting in single-pair modules achieving 2.2$''$ HPD with full illumination with 4.5 keV X-rays.  Every element of the technology, including mirror fabrication, coating, alignment, and bonding, continue to be improved to meet AXIS's requirements.We expect this technology will be at TRL-6 for AXIS by 2024.

\subsubsection{Detector}

The focal plane detector will be a CCD or related device, qualitatively similar to those flown on Chandra, Suzaku, and other recent X-ray observatories, but taking advantage of recent technical improvements. The 24$'$ field of view requires a 7.3 cm x 7.3 cm detector, or a mosaic of a small number of devices. The key technical challenges are pixel size, readout rate, and sufficiently low readout noise to ensure good low energy response. For our baseline detector layout, we will obtain $\sim 0.3''$ over the central 14$' \times 14'$ and 0.4$''$ over the entire 24$' \times 24'$ field of view.

With a plate scale of $\sim$ 46 $\mu$m arcsec$^{-1}$, in order to adequately sample the beam, the effective detector pixels should be $\sim$8 $\mu$m (0.17$''$).  Charge cloud diffusion in a suitably thick device ($\sim$100 $\mu$m depletion) allows centroiding of the incoming photon on sub-pixels scales and somewhat larger pixels of $\sim$16 $\mu$m (0.34$''$) can be used \cite{Bautz18, Falcone18}.

Higher readout rates and smaller pixels will substantially reduce the particle background through superior multi-pixel filtering of the larger charge clouds produced by cosmic rays and significantly reduce pile-up from bright point sources.  Current CCD technology achieves 2.5 Mpix s$^{-1}$ transfer rates with readout noise less than 7e- giving good spectral resolution and low energy capability.  A single 3cm x 3cm CCD with 8 $\mu$m pixels and 8 readout nodes (the most conservative case), has an integration time of 50 msecs, much shorter than the Chandra ACIS. By using multiple CCDs and additional readout nodes, the integration time can be reduced by a factor of at least 10. 

Backside-illuminated (BI) CCDs as flown on Chandra and Suzaku have good QE over the entire AXIS energy range, and NASA R\&A programs are underway to minimize the thickness of the filters required to block unwanted visible/UV radiation to give higher low energy throughput than Suzaku.  Present optical blocking filters give a combined total QE of 25\% at 0.2 keV, 75\% at 0.5 keV, and greater than 90\% above 1 keV.  A QE of 90\% up to 10 keV is achievable as demonstrated with the SXI CCDs aboard Hitomi \cite{Tsunemi16} , although this requires deep depletion that produces more lateral charge diffusion and can degrade spatial resolution and soft spectral response \cite{Miller18}.

\subsection{Mission Design}

We use the mission design parameters as developed by the GSFC Mission and Instrument Design Labs for the AXIS conceptual mission design. The most severe system requirement arises from the high angular resolution: the attitude must be reconstructed (not controlled) to $\sim$0.1$''$.  Only modest ($\sim$0.5$'$) pointing accuracy is needed, but maintaining the needed knowledge accuracy requires a stable thermal design), plus a focus mechanism and metrology system to monitor and remove structural variation. The key requirement is jitter, which in turn is driven by the CCD readout rate (the faster the readout rate, the more relaxed the jitter requirement). Our study has concluded that the best orbit for AXIS is a quasi-equatorial low earth orbit ($\sim$600 km) with as low an inclination as allowed by a Falcon 9 launch, given reasonable mass margins. Our spacecraft design allows for a high slew rate that will enable target of opportunity science similar to Swift as well as a high (~75\%) observing efficiency.


\begin{table}[h]
\label{tab:fonts}
\begin{center}       
\begin{tabular}{|l|l|} 
\hline
\rule[-1ex]{0pt}{3.5ex}  Mission Lifetime (yrs) & 5 \\
\hline
\rule[-1ex]{0pt}{3.5ex}  Orbit & LEO equatorial   \\
\hline
\rule[-1ex]{0pt}{3.5ex}  Mass (kg) & 1913  \\
\hline
\rule[-1ex]{0pt}{3.5ex}  Average power (W) & 1028  \\
\hline
\rule[-1ex]{0pt}{3.5ex}  Average/peak data rate (kbps) & 50  \\
\hline
\rule[-1ex]{0pt}{3.5ex}  Required pointing accuracy (arcmin) & 0.5 \\
\hline
\rule[-1ex]{0pt}{3.5ex}  Pointing knowledge (arcsec) & $\sim 0.1$  \\
\hline
\rule[-1ex]{0pt}{3.5ex}  Launch vehicle & Falcon 9 \\
\hline
\rule[-1ex]{0pt}{3.5ex}  Margin to LEO & $\sim 170$\%  \\
\hline
\end{tabular}
\caption{AXIS Mission Parameters from MDL} 
\end{center}
\end{table} 

\subsection{TRL Estimates}

The X-ray mirror has the lowest TRL. The segmented Si technology is at TRL 5 for a $\sim$5$''$ angular resolution system.  CCDs with 8 $\mu$m pixel size are at TRL 6. The baseline spacecraft consists entirely of heritage components (TRL $>$ 6); no technology development is needed.

\subsection{Cost Estimate}

The AXIS mission cost has been estimated  at the GSFC IDL and MDL, by applying PRICE-H to the MDL spacecraft design, instrument, and mirror.  For the costing of the instruments, mirrors, and spacecraft components, a minimum TRL of 6 was assumed.  Standard "wrap" factors were applied to estimate cost for Project Management (WBS 1.0), Mission Systems Engineering (2.0), Safety and Mission Assurance (3.0), and Mission Integration and Test (10.0).  Costs for Science (WBS 4.0), Mission Operations (7.0), and Ground System Development (9.0) were applied. The cost reserve was calculated by applying 30 percent to all WBS elements except Science (10 percent) and Launch Vehicle (0\%). The total cost including reserve, comes well below the nominal Probe maximum of \$1B.

Like all the Probe studies, the Final Report, due in Dec 2018 will compile the science case, the requirements flowdown, and the resulting mission design in preparation for input to the decadal survey.


\bibliography{report}   
\bibliographystyle{spiebib}   

\end{document}